\documentclass[12pt]{iopart}

\usepackage{cite}
\usepackage{setstack}
\usepackage[pdftex]{graphicx}
\usepackage[caption=false,font=footnotesize]{subfig}
\usepackage{siunitx}
\usepackage{placeins}
\usepackage{float}
\usepackage{hyperref}
\usepackage{color}
\DeclareGraphicsExtensions{{.pdf},{.png}}
\graphicspath{{figures/}}

\newcommand\chem[1]{\ensuremath{\mathrm{#1}}}

\begin{document}

\title[MC simulations of spin transport in nanoscale \chem{In_{0.7}Ga_{0.3}As} transistors]{Monte Carlo simulations of spin transport in nanoscale \chem{In_{0.7}Ga_{0.3}As} transistors: temperature and size effects}

\author{B.~Thorpe$^1$, S.~Schirmer$^2$, and K.~Kalna$^1$}

\address{$^1$ Swansea Academy of Advanced Computing, Swansea University, Bay Campus, Swansea, SA1 8EN, United Kingdom\\ $^2$ Department of Physics, Faculty of Science \& Engineering, Swansea University, Singleton Park, Swansea, SA2 8PP, United Kingdom\\ $^3$ Nanoelectronic Devices Computational Group, Department of Electronic and Electrical Engineering, Faculty of Science \& Engineering, Swansea University, Bay Campus, Swansea, SA1 8EN, United Kingdom}

\ead{b.j.thorpe@swansea.ac.uk, s.schirmer@swansea.ac.uk, k.kalna@swansea.ac.uk}

\begin{abstract}
  Spin-based metal-oxide-semiconductor field-effect transistors (MOSFET) with a high-mobility III-V channel are studied using self-consistent quantum corrected ensemble Monte Carlo device simulations of charge and spin transport. The simulations including spin-orbit coupling mechanisms (Dresselhaus and Rashba coupling) examine the electron spin transport in the \SI{25}{nm} gate length \chem{In_{0.7}Ga_{0.3}As} MOSFET. The transistor lateral dimensions (the gate length, the source-to-gate, and the gate-to-drain spacers) are increased to investigate the spin-dependent drain current modulation induced by the gate from room temperature of \SI{300}{K} down to \SI{77}{K}.  This modulation increases with increasing temperature due to increased Rashba coupling.  Finally, an increase of up to $20$~nm in the gate length, source-to-gate, or the gate-to-drain spacers increases the spin polarization and enhances the spin-dependent drain current modulation at the drain due to polarization-refocusing effects.
\end{abstract}

\vspace{2pc}
\noindent{\it Keywords}:
InGaAs FET, spin transport, Dresselhaus and Rashba coupling, Monte Carlo simulation.

\maketitle

\section{\label{sec:intro}Introduction}

Transistors using electron spin in their operation have been proposed as an alternative to conventional devices for a number of years to unlock potential novel functionality, reduce power consumption, and increase performance~\cite{Awschalom2013,Wolf2001}. Among the most promising spin-based semiconductor devices is the spin field-effect transistor (spinFET)~\cite{Datta1990}, a candidate for future high-performance digital computing and memory with ultra low energy needs~\cite{IRDS2020}. The spinFET architecture is similar to that of a conventional semiconductor transistor, with the key difference that the source and drain contacts are ferromagnetic. The source injects spin polarised carriers into the transistor channel. The drain, by contrast, acts as a spin filter by preferentially transmitting carriers, whose magnetic moments align with that of the source~\cite{Moodera2007}.  A source-to-drain current can be then modulated by the voltage applied to a gate contact due to a spin rotation caused by the Rashba spin-orbit coupling~\cite{Rashba1959}. This model was generalized to include other effects such as spin dephasing due to bulk inversion asymmetry (so called Dresselhaus coupling)~\cite{Bournel2000} and used to predict spin transport effects for 2D III-V heterostructure HEMTs~\cite{Bournel2001, MinShen2004}.

Spin dephasing is one of the main limiting factors for a spinFET \cite{Schliemann2003, Cartoixa2003}. To minimise the spin dephasing and maximise the spin control \cite{Scherubl2016}, a tuning of Rashba coupling was independently suggested~\cite{Schliemann2003, Cartoixa2003}. The strength of the Rashba coupling, which depends on the electric field, can be tuned by applying a gate voltage. In theory, we can adjust the gate voltage so that the Rashba coupling effectively cancels the Dresselhaus coupling. However, this approach comes with a significant drawback. The gate bias required to cancel the spin-orbit coupling results in an off-current half of the on-current~\cite{Bandyopadhyay2004}, making such devices unsuitable for digital applications because the off-current must be orders of magnitude smaller (typically $5$ to $8$ orders of magnitude)~\cite{TajalliTCS2011}. The simulation data suggests that the on-off ratio would only be around 15-20\% in a real device~\cite{E.Shafir2004}. Therefore, this work will not consider the transistors based on a tuning of Rashba coupling but will focus on inversion channel FETs~\cite{Sugahara2016}. 

Finally, the FET studied has a \chem{In_{0.7}Ga_{0.3}As} channel because the control of spin-orbit interaction by a gate voltage was reported experimentally in \chem{In_{0.53}Ga_{0.47}As}/\chem{In_{0.52}Al_{0.48}As} and \chem{InP}/\chem{In_{0.77}Ga_{0.23}As}/\chem{InP} heterostructures \cite{Nitta:1997, Engels:1997}. The FETs with a channel made of InGaAs are intensively studied as potential replacements for Si channels for More Moore digital solutions \cite{Benbakhi2012}. The recent advances in the nanoscale InGaAs FETs include gate-all-around nanowire FETs \cite{Kilpi:2021, Tomioka:2021}, gate-all-around nanosheet FETs \cite{Lee:2018}, and FinFETs \cite{Lu:2018}.

In this work, a new 2D (real space) electron-spin quantum corrected finite-element (FE) ensemble Monte Carlo (MC) device simulation tool~\cite{Thorpe2017} is employed to investigate the effects of lattice temperature and device dimensions in a \SI{25}{nm} gate length \chem{In_{0.7}Ga_{0.3}As} spinFET. The ensemble MC device technique is a semi-classical transport technique for semiconductor devices~\cite{C.Jacoboni1989} self-consistently simulating the carrier movement classically and the carrier scattering quantum-mechanically coupled with solutions of Poisson equation (accounting for long-range electron-electron interactions) in a device simulation domain. The 2D MC device simulation tool~\cite{Kalna2007} i) has incorporated quantum corrections~\cite{Kalna2008, N.Seoane2016} using the effective quantum potential~\cite{Ferry2000}, and ii) has been enhanced to include a non-equilibrium spin transport~\cite{Thorpe2017} so is capable to accurately model a highly non-equilibrium electron-spin transport at nanoscale~\cite{AynulIslam2011}.

An electron spinFET brings many advantages into spin based logic operations when compared to the current electron charge based logic. The speed of spin logic operation would be substantially faster, even the increase depends on the efficiency of a spin injection at the source, operational temperature of the spinFET, and on the efficiency of spin detection in the drain. A power dissipation of spin logic operations would be orders of magnitude smaller than a power dissipation of charged electron logic operations. A spinFET would also not suffer from any short channel effects when further scaled down with a tiny off-spin-current, a non-existent DIBL, and a sub-threshold slope below 60 mV/dec at 300 K. The overall area of a spinFET would be much smaller than the area of a classical FET allowing for a substantial increase in a spin transistor density on a chip. Finally, a spinFET would be able to operate in a quantum logic, not just in classical logic operations.

This paper is organized as follows. In Sec.~\ref{sec:Design}, the device design and geometry, and the simulation technique are described. In Sec.~\ref{sec:Temperature}, the underlying theory for the effects of lattice temperature on spin-orbit coupling is summarized, and the simulation results are discussed. In Sec.~\ref{sec:Geometry}, the effects of device dimensions on spin transport are explored by varying the length of the gate, and the length of the left and right spacers, respectively. Conclusions are drawn in Sec. \ref{sec:Conclude}. Finally, two appendices (\ref{AppA} and \ref{AppB}) collect the parameters used for temperature dependence of band energies, and show a spin polarization angle and a magnitude of the spin polarization along the device channel, respectively.

\section{\label{sec:Design}Device Design and Simulation}
\subsection{\label{sec:transistor}Nanoscale Transistor Geometry}

\begin{figure}
\centering
\includegraphics[width=\columnwidth]{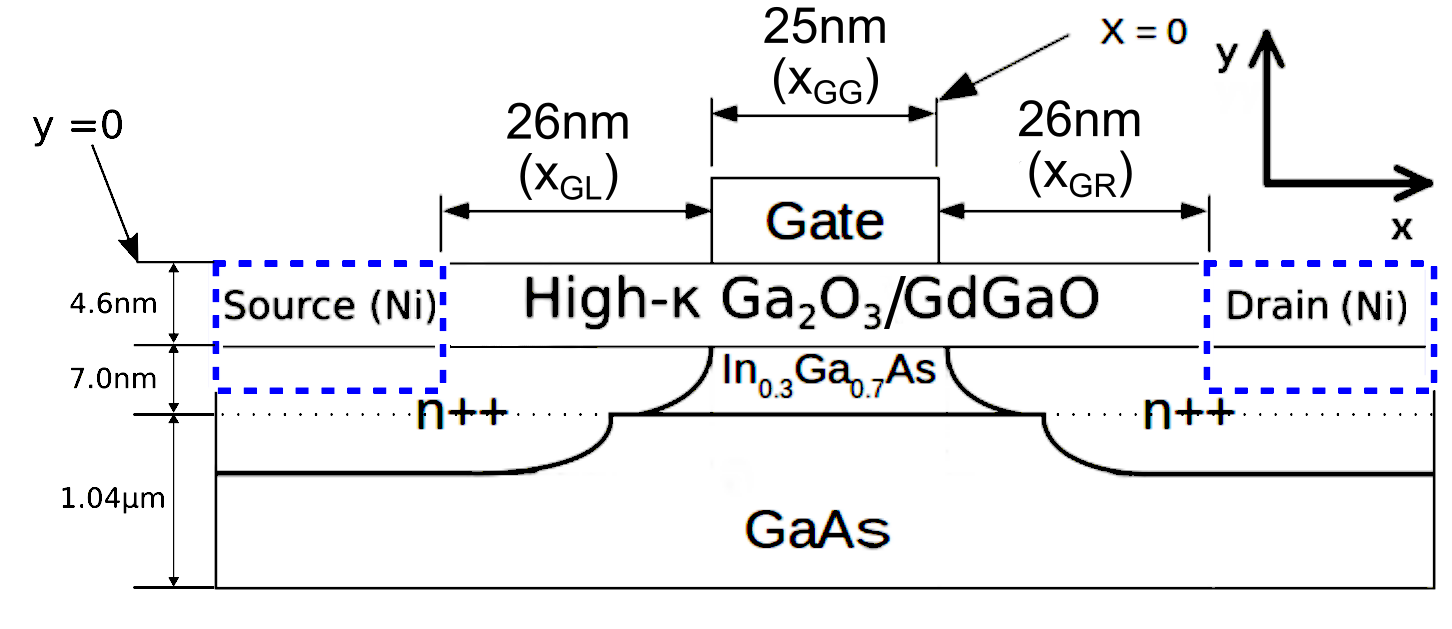}
\caption{Cross-section of $n$-channel \chem{In_{0.3} Ga_{0.7}As} MOSFET with a gate length of \SI{25}{\nano\metre} and indicated lateral and transverse dimensions. The dielectric layer with a thickness of $4.6$~nm between the channel and the gate is made of a gallium oxide/gadolinium gallium oxide stack (\chem{Ga_{2}O_{3}/GdGaO}), forming a high-$\kappa$ dielectric layer with a low density of interface states \cite{Kalna2008}. The dashed blue lines indicate the position of the source and the drain contacts, which are mimicked by electron reservoirs in the simulations. The labels $X_{GL}$,$X_{GG}$ and $X_{GR}$ refer to the source-to-gate spacer, the gate length and the gate-to-drain spacer, respectively.} \label{fig:diagram}
\end{figure}

A nanoscale \chem{In_{0.7}Ga_{0.3}As} FET, which has been designed as a potential contender to \chem{Si} metal-oxide-semiconductor (MOS) FET for the sub-$10$~nm planar low-power technology~\cite{AynulIslam2011, Skotnicki2005}, is specifically investigated to study the spin transport in a device structure which can be realistically fabricated within the complementary MOS (CMOS) technology \cite{delAlamo2011}. A schematic of the device is given in Fig.~\ref{fig:diagram}. The transistor is fabricated on a \chem{Si} substrate by growing a \SI{400}{\nano\metre} \chem{GaAs} buffer layer (using intermediate layers or graded layers or a wafer bonding), a \SI{7}{\nano\metre} thick \chem{In_{0.3}Ga_{0.7}As} channel grown epitaxially, a \SI{4.6}{\nano\metre} layer of high-$\kappa$ \chem{Ga_{2}O_{3}}/(\chem{Gd_{x}Ga_{1-x}})(\chem{GGO}, $\kappa=20$) separating the channel from a metal gate with a work function of \SI{4.05}{\electronvolt}. The \chem{GaAs} buffer has a background uniform $p$-type doping of \SI{1e18}{\per\cubic\centi\metre}. The source/drain (S/D) in the transistor has a $n$-type Gaussian-like doping including doping extensions, as depicted in Figure \ref{fig:diagram}, with a maximum doping of \SI{2e19}{\per\cubic\centi\metre}.

A spin transistor requires source and drain contacts that are ferromagnetic, the source acting as a spin injector, the drain contact as a spin detector.  Since we are primarily interested in the spin transport along the transistor channel, the spin-injection from the source is assumed to be $100$\% efficient.  The idealistic $100$\% injection efficiency is often used in spin transport simulations~\cite{Bournel1997,Bournel1998,Bournel2000,Bournel2001} because the injection efficiency depends on particular details of ferromagnetic contact fabrication.  The idealistic $100$\% injection efficiency of the source can be relatively easily adjusted to more realistic injection efficiencies~\cite{Thorpe2017}.  Highly efficient electrical spin injection has been demonstrated for \chem{AlGaAs} 2DEGs using Mn doped Esaki diodes with reported efficiencies as high as 75\% \cite{Oltscher2014} albeit at extremely low temperatures (less than \SI{4}{\kelvin}).

\subsection{Spin Transport Simulation}

The transistor simulations are performed in the device on-region (providing the logical $1$ in a binary logic) at a source-drain voltage of \SI{0.9}{\volt} and a gate voltage of \SI{0.7}{\volt} \cite{AynulIslam2011}.  The MC simulations are run with $100,000$ super-particles in a time step of \SI{1}{\femto\second} (to minimise plasma oscillations) for a total time of \SI{10}{\pico\second}.  Three injection spin directions are considered with spins aligned parallel to the direction of transport ($x$ axis), the growth direction of the heterostructure ($y$ axis), and the plane of the 2DEG ($z$ axis).  The spins of the electrons in the channel are initially assumed to be randomly oriented such that there is no net magnetic field.  For each time-step, the average spin polarization vector of the current $\mathbf{S}(t)$ is recorded at the drain contact.  The average components of the spin polarization vector are obtained by averaging the individual $x$, $y$ and $z$ components of the spin polarization vectors of all electrons located at different positions along the channel, in particular at the (left) edge of the drain at the time $t$.  The magnitude of $\mathbf{S}(t) \leq 1$ defines the polarization in the direction of $\mathbf{S}(t)$, with $1$ corresponding to the maximum spin polarization.  This provides a measure of the loss of spin polarisation due to scattering as the electrons traverse the channel. 

The angle $\theta$ between $\mathbf{S}_{\mathrm{drain}}$ and the injected state $\mathbf{S}_{\mathrm{inj}}$ is given by
\begin{equation*}
  \theta = \cos^{-1}\left(\frac{\left|{\mathbf{S}_{\mathrm{drain}}}\right|}{\mathbf{S}_{\mathrm{drain}}\cdot\mathbf{S}_{\mathrm{inj}} }\right).
\end{equation*}
The magnitude and the angle of the polarization vector determine the expected modulation of the drain current, $ \xi(M_D,\theta)$ (Eq.~(9) \cite{Thorpe2017}) as
\begin{equation}
  \xi(M_D,\theta) = \frac{1+M_D\cos\theta}{1+M_D}
\end{equation}
where $M_D = |\mathbf{S}_{\rm drain}|$ is the magnitude of the polarization vector at the drain edge.  For $\theta=0$, we have $\xi(M_D,\theta)=1$.  This motivates the definition of the spin-dependent drain current modulation, $V_S$, as
\begin{equation}\label{vis}
   V_S(M_D,\theta) = 1 - \xi(M_D,\theta) = \frac{M_D}{M_D+1} (1-\cos\theta)
\end{equation}
% 1 - (1+M -M + M cos)/(1+M) = - (-M+Mcos)(1+M) = M/(M+1)*(1-cos)
in the following referred to simply as the spin modulation. $V_S$ ranges from $0$ for $\theta=0$ to $1$ for $\theta=180^\circ$ and $M_D=1$ and is a measure distinguishing the on- and off-state from the drain current.  Maximum modulation requires a high spin polarization and a large effective rotation angle $\theta$.

In the following, when we refer to the (spin) polarization angle, we shall imply the angle $\theta$ of the polarization vector at a given time or point in space relative to the initial polarization vector of the injected spins, while (spin) polarization shall refer to the magnitude of the (spin) polarization vector.  Spin polarization always refers to the electron spins, and we shall often simply use polarization.

\section{Temperature Dependence}\label{sec:Temperature}

Our first aim is to explore the temperature dependence of the spin polarization at the drain edge for a fixed geometry and at a fixed gate voltage.  The former depends primarily on the temperature dependence of the Rashba and Dresselhaus coupling. All characteristics of the spin transistor (spin polarization, spin angle) are monitored by averaging physical quantities of interest over the all particles and a simulation time as usual in the Monte Carlo technique \cite{C.Jacoboni1989}, with the temperature dependence taken into account through the change in energy assuming a Fermi-Dirac distribution \cite{AynulIslam2011}.

\subsection{Rashba and Dresselhaus coupling}

\begin{figure}[t]
\subfloat[Dresselhaus coupling]{\includegraphics[width=0.48\textwidth]{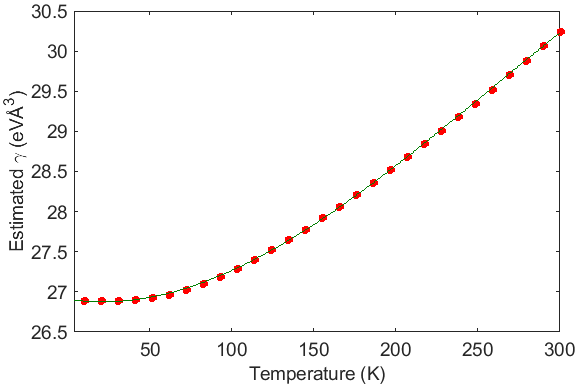}} \hfill
\subfloat[Rashba coupling]{\includegraphics[width=0.48\textwidth]{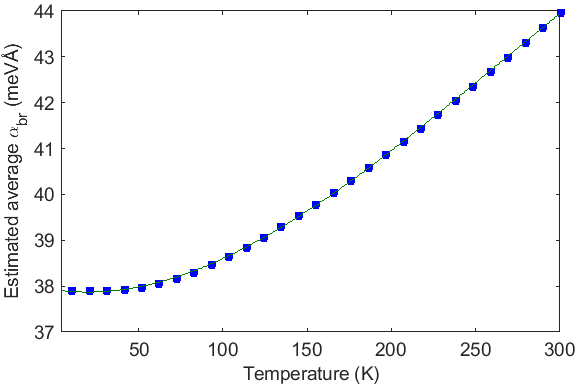}}
\caption{Temperature dependence of Dresselhaus ($\gamma$) and Rashba ($\alpha_{br}$) coefficients in \chem{In_{0.7}Ga_{0.3}As} between \SI{4}{\kelvin} and \SI{300}{\kelvin}.  The Rashba coefficient is estimated using an average electric field of \SI{4.219E7}{\volt\per\metre} present at $V_G=\SI{0.7}{\volt}$ and $V_D=\SI{0.5}{\volt}$.  The dark green lines are cubic fits to the data.  The shape of the Rashba curve agrees well with similar calculations for \chem{GaAs} \cite{J.Fabian2007}.} \label{fig:SOC-temp}
\end{figure}

The temperature dependence of Dresselhaus and Rashba parameters $\gamma$ and $\alpha_{br}$ relates to the temperature dependence of the lattice constant $a_0$, leading to an increase in the band energies $E_0,E_1,\Delta_0$ and $\Delta_1$ with decreasing temperature.  This effect has been studied for $\chem{GaAs}$~\cite{Wang2016,Hubner2009}, and the temperature dependence of the bandgap energies for \chem{GaAs} has been investigated experimentally~\cite{Hubner2009, Lautenschlager1987}, leading to the relation
\begin{equation}\label{bandtemp}
E_g(T_L) = E_{g0}-\alpha_B\bigg(1+\frac{2}{\exp(\theta/T_L)-1}\bigg),
\end{equation}
where $T_L$ is the lattice temperature and the parameters $E_{g0},\alpha_B$ and $\theta$ are obtained by fitting experimental data. The temperature dependence for \chem{InAs}, needed to fit a \chem{In_xGa_{1-x}As} ternary compound, with $x$ being the content of \chem{In}, is not as well documented in the literature, with only limited experimental data for the fitting parameters $\theta$ and $\alpha_B$ published for the inter-band energies~\cite{Kim2012, Passler2001}. The fitting parameters used to calculate the band energy dependence in our simulations are listed in Table~\ref{temp-param} in ~\ref{AppA}. The resulting dependence of the Dresselhaus ($\gamma$) and Rashba ($\alpha_{br}$) coefficients as a function of the lattice temperature is shown in Fig.~\ref{fig:SOC-temp}. Both coefficients increase non-linearly due to the changes in the interband energies, both varying cubically as expected. The shape of the Rashba curve fits well with similar calculations for \chem{GaAs} \cite{J.Fabian2007}.

\subsection{Polarization at Drain Edge}

\begin{figure*}
\subfloat[Spin polarization at the drain edge.\label{fig:MagVsTemp}] {\includegraphics[width=0.32\textwidth]{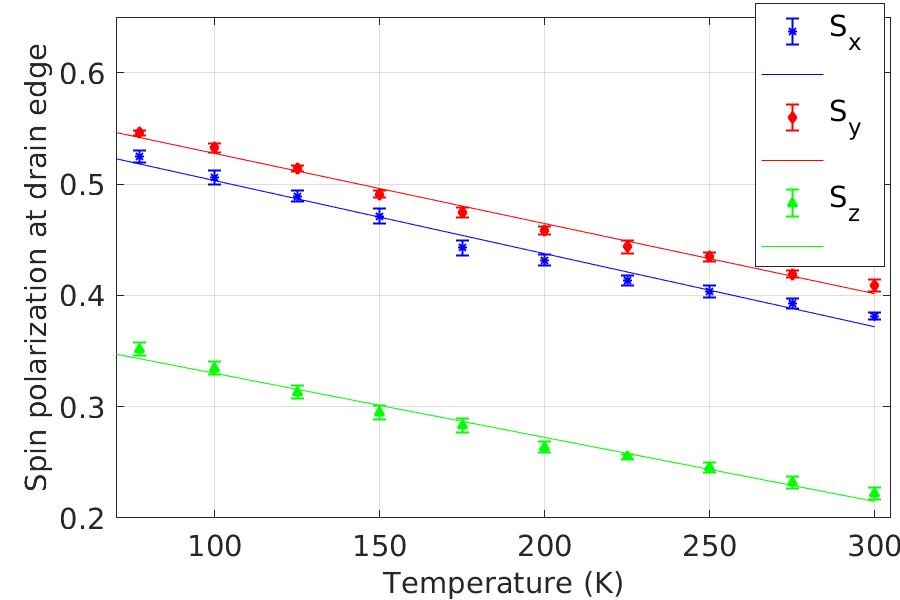}} \hfill
\subfloat[Polarization angle $\theta$ at the drain edge.\label{fig:AngleVsTemp}] {\includegraphics[width=0.32\textwidth]{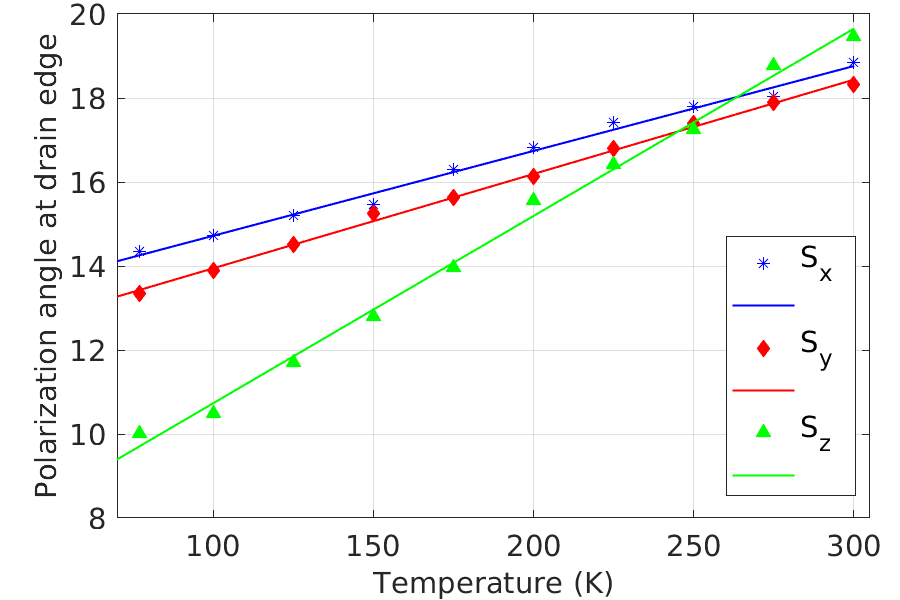}} \hfill
\subfloat[Spin modulation at the drain edge. \label{fig:T-dependence}]
{\includegraphics[width=0.32\textwidth]{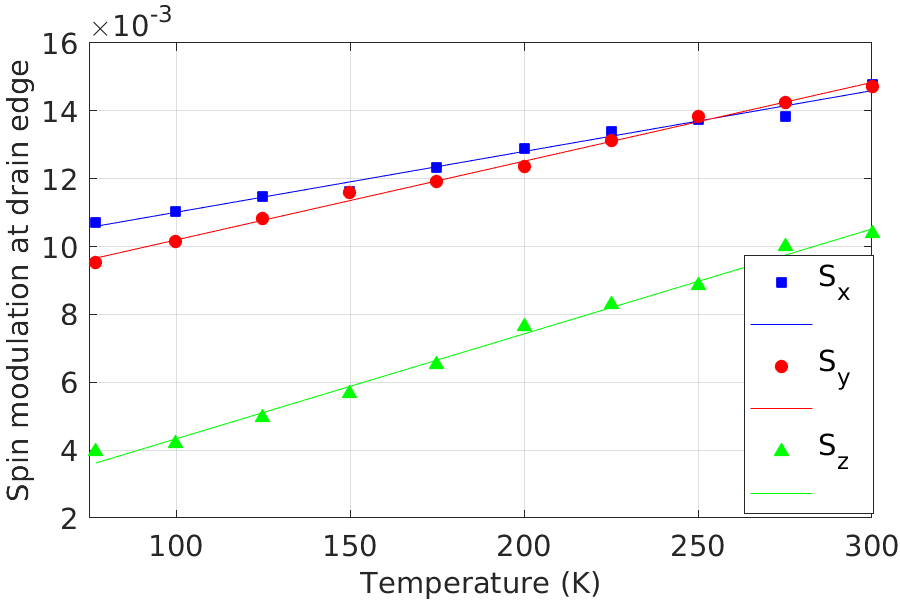}}
\caption{Lattice temperature dependence of the spin polarisation, the spin polarisation angle $\theta$, and the spin modulation, all at the drain edge at $V_G$=\SI{0.7}{\volt} and $V_D$=\SI{0.9}{\volt}. The linear fits elucidate trends.}
\end{figure*}

The spin dynamics are explored between \SI{77}{\kelvin} and \SI{300}{\kelvin} at a gate voltage ($V_G$) of \SI{0.7}{\volt} and a drain voltage ($V_D$) of \SI{0.9}{\volt} (the transistor drive bias \cite{delAlamo2011}). Note that the \chem{In_{0.7}Ga_{0.3}As} FET has a threshold voltage ($V_T$) of 0.2 V as required to provide $V_D-V_T\equiv V_G$=0.7~V \cite{Kalna2007, Kalna2008}. This covers practical temperatures from liquid nitrogen cooling up to room temperature, in which the MC simulator using the Fermi-Dirac statistics has been tested \cite{IslamSST2011}. Fig.~\ref{fig:MagVsTemp} shows that the spin polarization across the device channel increases linearly with decreasing temperature. The lowering of the temperature decreases the number of electron scattering events as the electrons travel through the channel, which is expected to decrease the decay of the spin polarization, therefore leading to a higher net spin polarization at the drain edge.  However, both the Rashba and Dresselhaus coupling also decrease with temperature, which reduces the rotation angle $\theta$ between the initial spin polarization and the spin polarization at the drain edge, as shown in Fig.~\ref{fig:AngleVsTemp}.  Indeed, for $S_z$ injection polarization, the rotation angle $\theta$ at \SI{300}{K} is about twice as large as for \SI{77}{K}.  Thus, we have two competing effects, increasing spin polarization and decreasing rotation angles.  Combining both effects shows that the latter is dominant, i.e., the spin modulation by the gate actually \emph{increases} with increasing temperature (Fig.~\ref{fig:T-dependence}) due to increasing spin-orbital Dresselhaus and Rashba coupling of electrons with increasing temperature. This spin-orbital coupling is induced by the high electric fringing field present beneath the gate, especially, at the drain side of the gate \cite{KalnaAsenov2002}.

\section{Device Scaling}\label{sec:Geometry}

\begin{figure*}[bt]
\subfloat[Spin polarization at the drain edge.\label{fig:MagVsLeft}] {\includegraphics[width=0.32\textwidth]{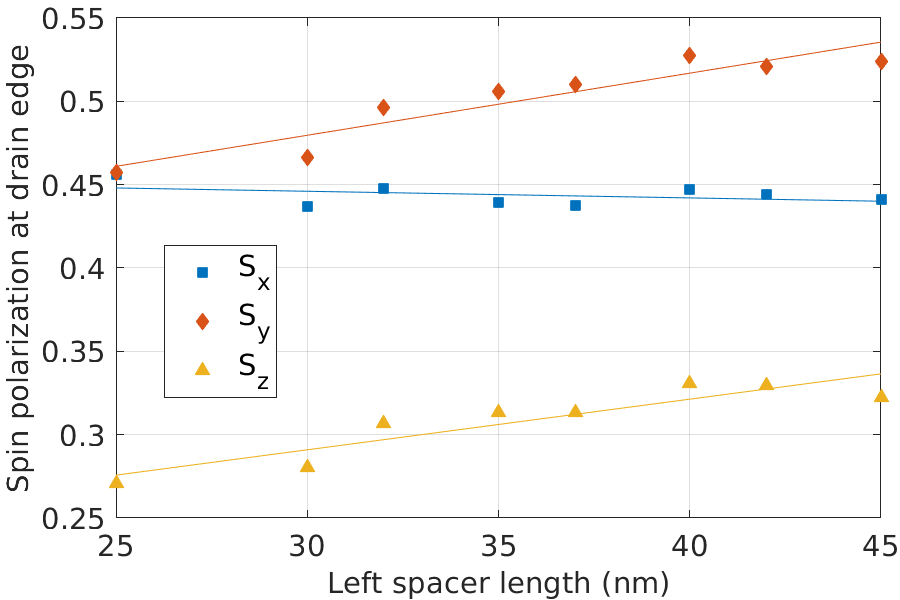}} \hfill
\subfloat[Polarization angle $\theta$ at the drain edge.\label{fig:thetaVsLeft}] {\includegraphics[width=0.32\textwidth]{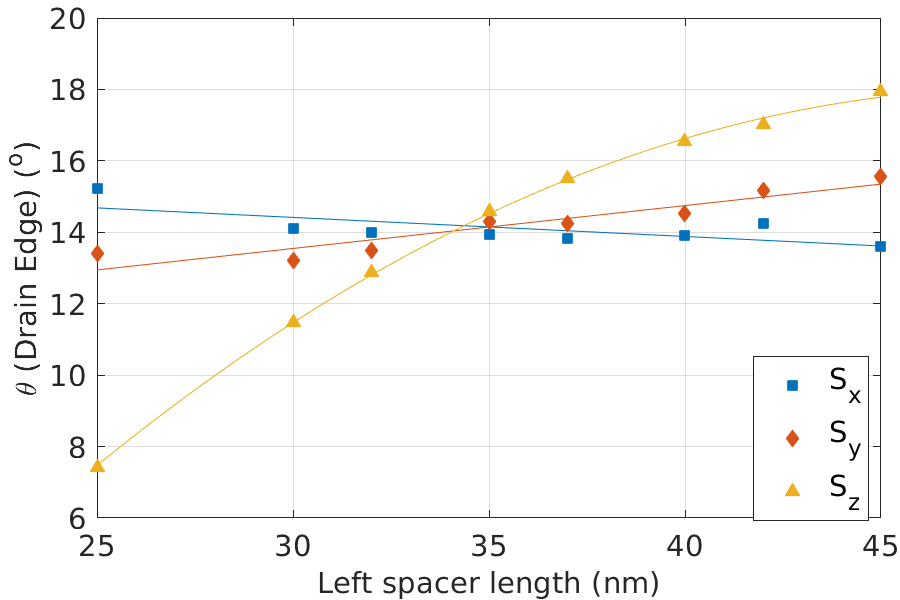}} \hfill
\subfloat[Spin modulation at the drain edge.\label{fig:VisLS}] {\includegraphics[width=0.32\textwidth]{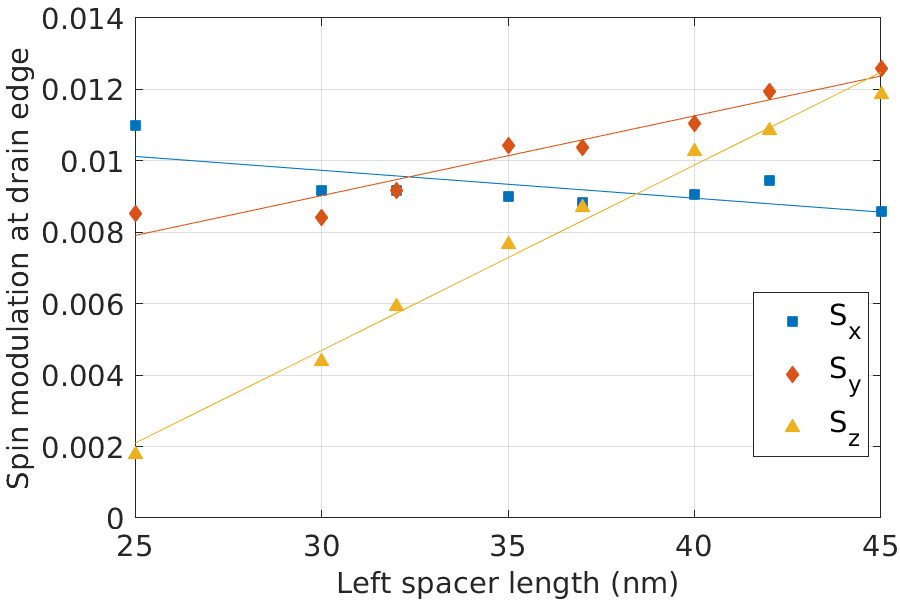}}
\\
\subfloat[Minimum spin polarization.\label{fig:MagMinVsLS}]
{\includegraphics[width=0.32\textwidth]{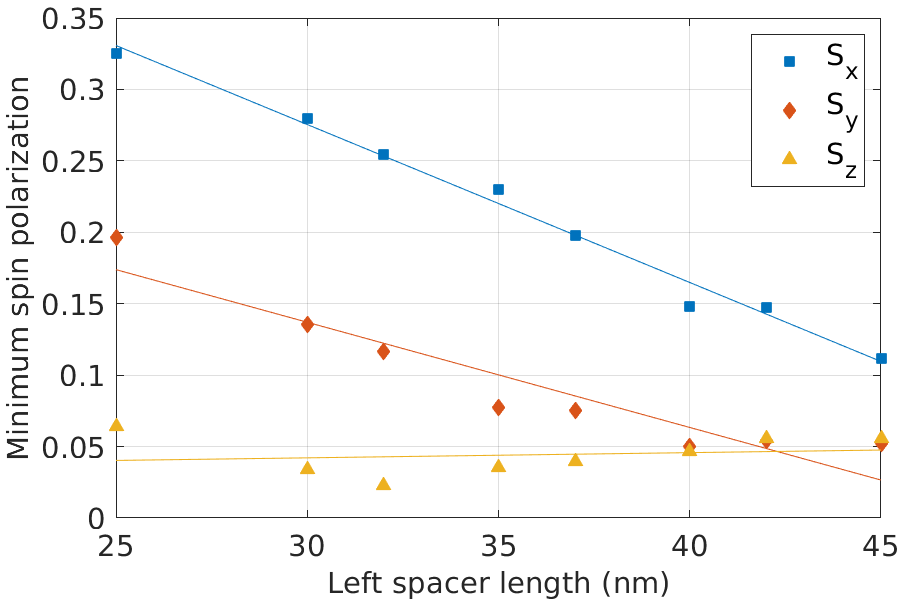}} \hfill
\subfloat[Spin polarization recovery.\label{fig:MagRecVsLS}] {\includegraphics[width=0.32\textwidth]{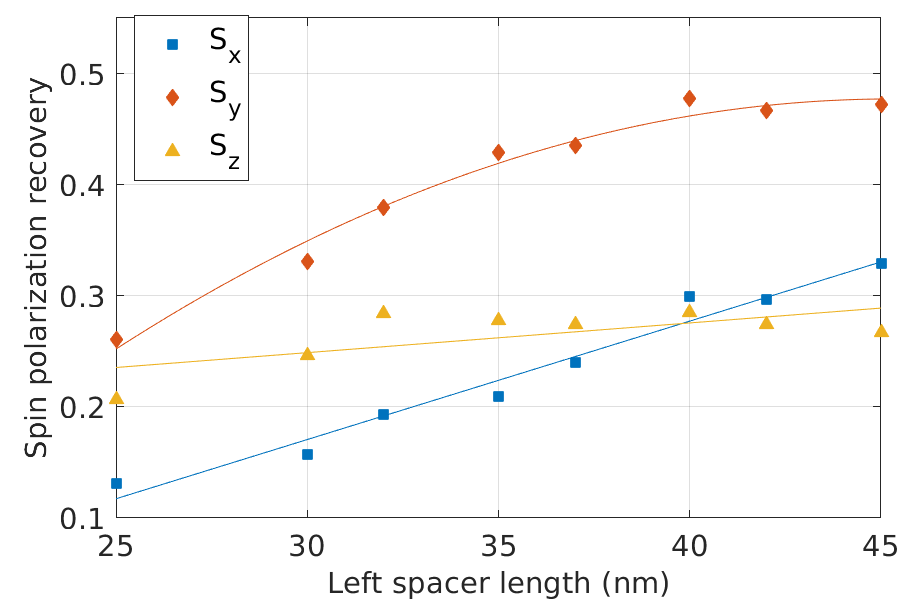}} \hfill
\subfloat[Maximum rotation angle $\theta$.\label{fig:AngMaxVsLS}] {\includegraphics[width=0.32\textwidth]{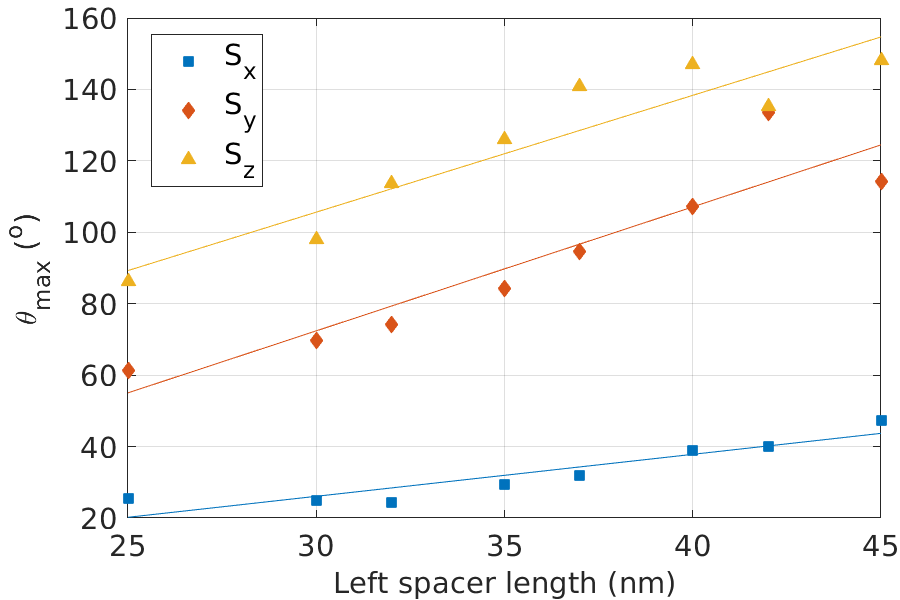}}
\caption{Spin polarization, rotation angle $\theta$, spin modulation [Eq. (\ref{vis})], all at the drain, then minimum spin polarization, spin polarization recovery, and maximum rotation angle $\theta_{\textrm{max}}$ in $x$, $y$, and $z$ directions (symbols) as a function of the source-to-gate spacer length ($V_G$=\SI{0.7}{\volt} and $V_D$=\SI{0.9}{\volt})} with fits to elucidate.
\label{fig:SourceScaling}
\end{figure*}

In this section, we aim to elucidate the effect of the device scaling on the drain edge polarization. Therefore, a spin polarization, a rotation angle $\theta$, and a spin modulation defined by Eq. (\ref{vis}) at the drain contact are plotted in Figs.~\ref{fig:GateScaling}-\ref{fig:SourceScaling} together with a minimum spin polarization, a spin polarization recovery, and a maximum rotation angle $\theta_{\textrm{max}}$ in $x$, $y$, and $z$ directions at operational bias of a gate voltage of \SI{0.7}{\volt} and a drain voltage of \SI{0.9}{\volt} (a so-called drive voltage). The spin polarization at the drain edge is an average spin polarization of all electrons entering the drain (modelled by a reservoir in the simulations) over all simulation time. All the following quantities are obtained also by averaging all electrons along the channel over all simulation time. The rotation angle of the polarization at the drain edge is a cumulative effect of the Rashba and Dresselhaus coupling experienced by electron spins as they move across the channel. The minimum spin polarization is the lowest spin polarization occurring along the channel, the spin polarization recovery is the difference between the spin polarization at the drain edge and the minimum of the spin polarization along the channel, and the maximum rotation angle $\theta_{\textrm{max}}$ is the maximum polarization angle reached along the channel.

\begin{figure*}[bt]
\subfloat[Spin polarization at the drain edge.\label{fig:MagVsRight}] {\includegraphics[width=0.32\textwidth]{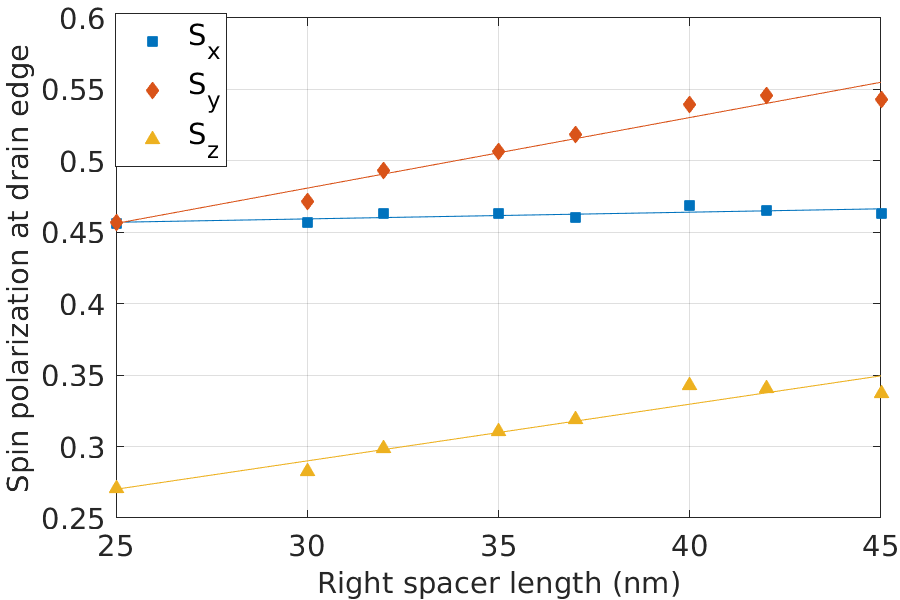}} \hfill
\subfloat[Polarization angle $\theta$ at the drain edge.\label{fig:thetaVsRight}]{\includegraphics[width=0.32\textwidth]{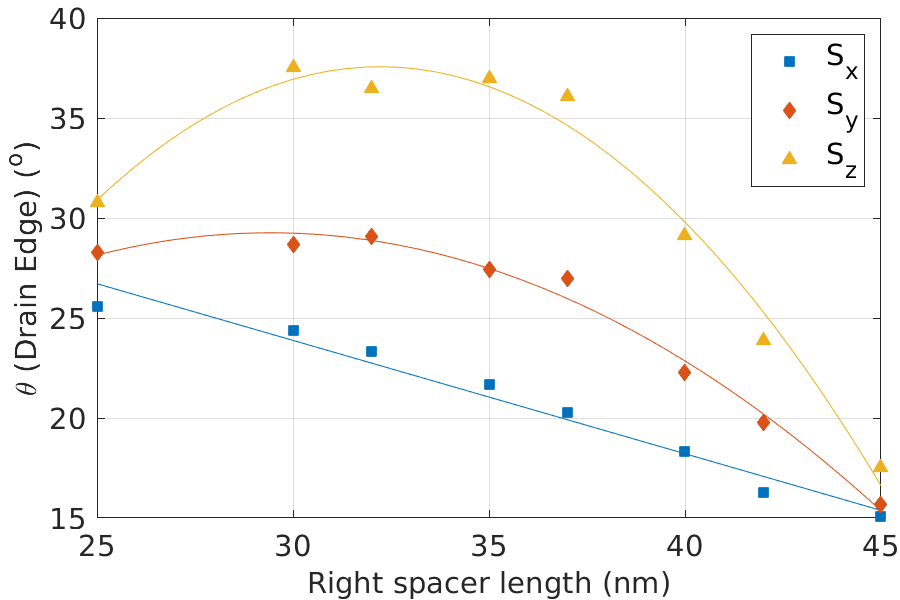}} \hfill
\subfloat[Spin modulation at the drain edge.\label{fig:VisRS}] {\includegraphics[width=0.32\textwidth]{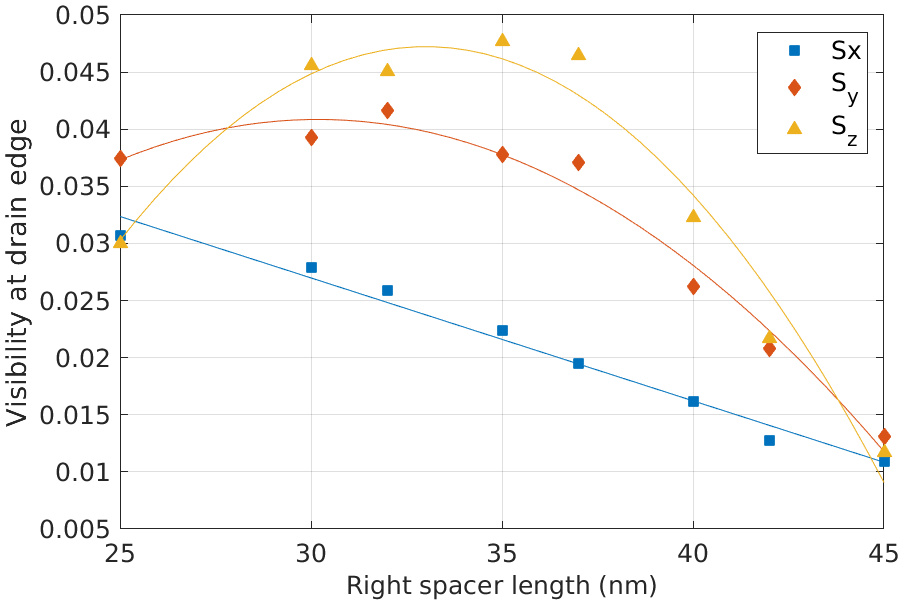}} 
\\
\subfloat[Minimum spin polarization.\label{fig:MagMinVsRS}]
{\includegraphics[width=0.32\textwidth]{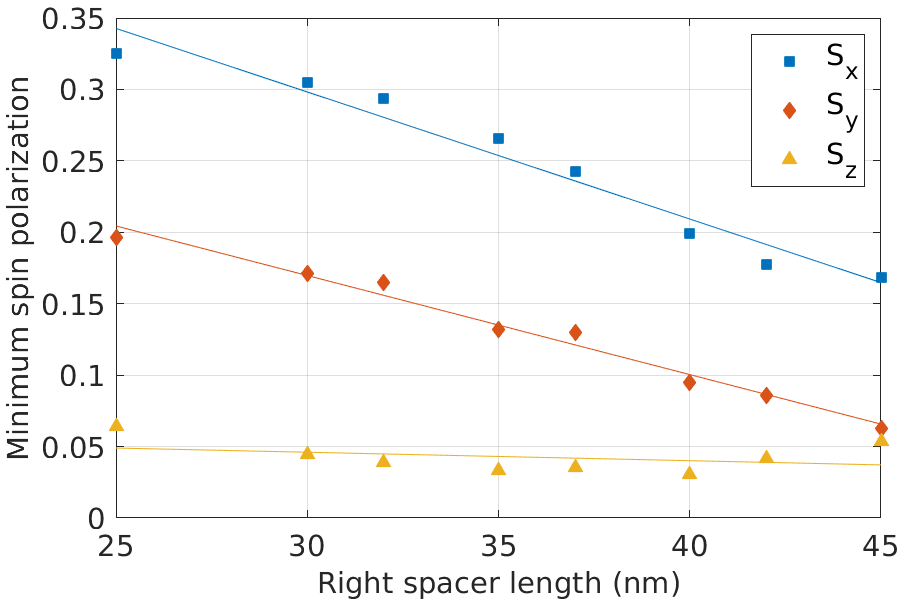}} \hfill
\subfloat[Spin polarization recovery.\label{fig:MagRecVsRS}] {\includegraphics[width=0.32\textwidth]{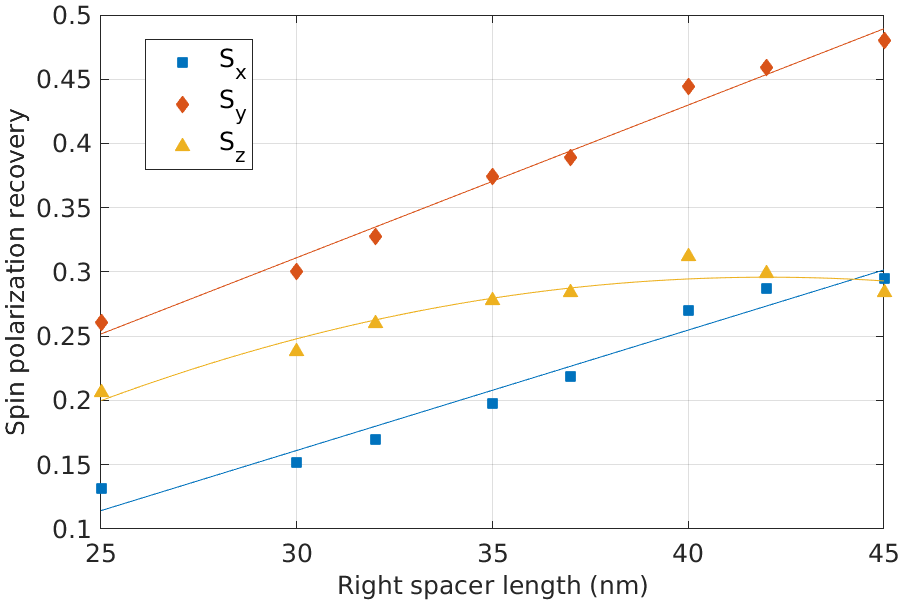}} \hfill
\subfloat[Maximum rotation angle $\theta$.\label{fig:AngMaxVsRS}] {\includegraphics[width=0.32\textwidth]{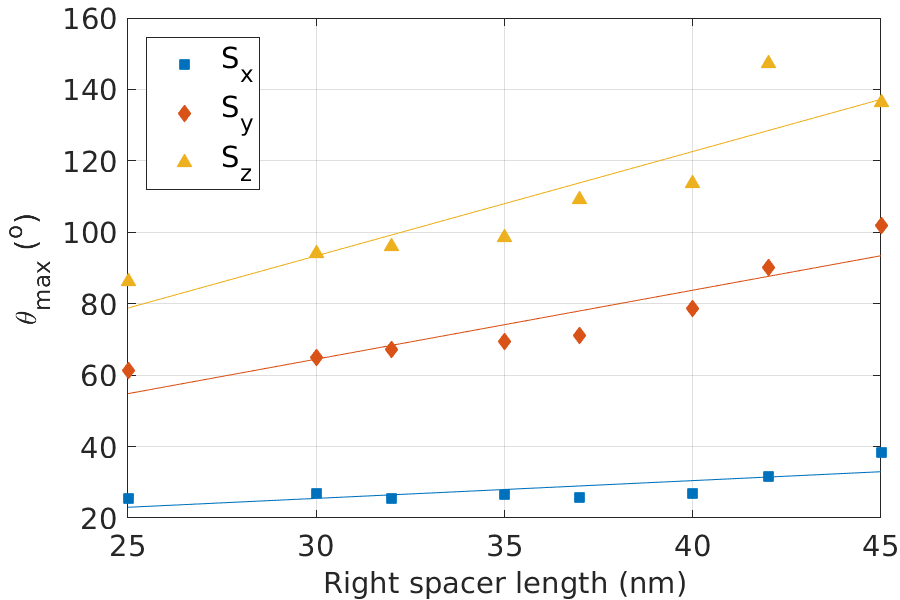}} 
\caption{Spin polarization, rotation angle $\theta$, spin modulation [Eq. (\ref{vis})], all at the drain, then minimum spin polarization, spin polarization recovery, and maximum rotation angle $\theta_{\textrm{max}}$ in $x$, $y$, and $z$ directions (symbols) as a function of the gate-to-drain spacer length ($V_G$=\SI{0.7}{\volt} and $V_D$=\SI{0.9}{\volt}) with linear fits to elucidate trends.}
\label{fig:DrainScaling}
\end{figure*}

If the channel length, or the distance between the gate and the source, or the distance between the gate and the drain (so-called spacers) is decreased/increased, the spatially distributed electric field will increase/decrease, affecting the spin-orbital coupling.  Therefore, we vary the three transistor design parameters: the length of the gate ($X_{GG}$) shown in Fig.~\ref{Fig:GateLength}, the distance between the gate and the source ($X_{GR}$) in Fig. \ref{fig:SourceScaling}, and the distance between the gate and the drain ($X_{GL}$) in Fig. \ref{fig:DrainScaling}. All other device dimensions and material parameters (e.g., the thickness of the layers in the $y$-direction and their composition) remains the same.

\subsection{Gate Length Dependence}\label{sec:Scaling}

\begin{figure*}[bt]
\subfloat[Spin polarization at the drain edge.\label{fig:MagVsGate}]
{\includegraphics[width=0.32\textwidth]{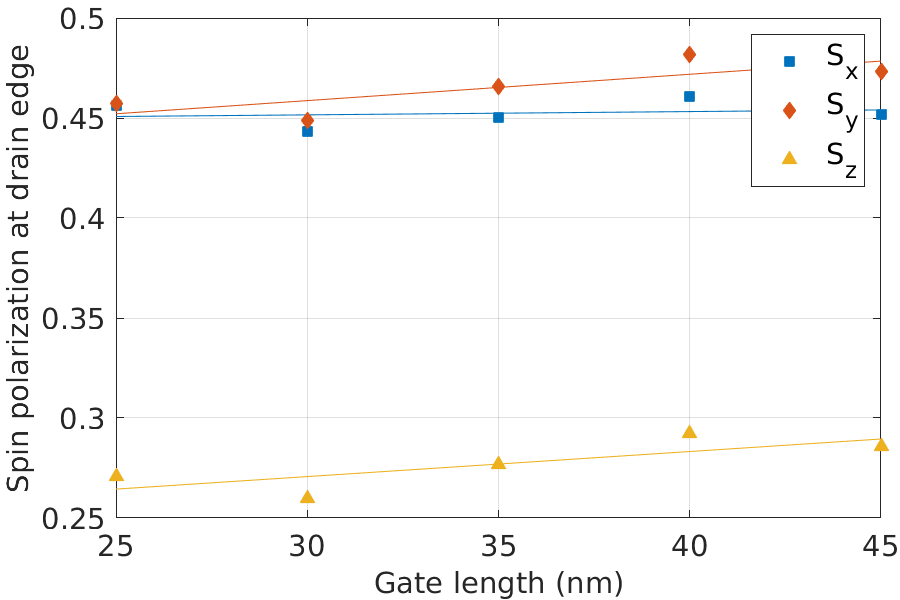}} \hfill
\subfloat[Polarization angle $\theta$ at the drain edge.\label{fig:thetaVsGate}] {\includegraphics[width=0.32\textwidth]{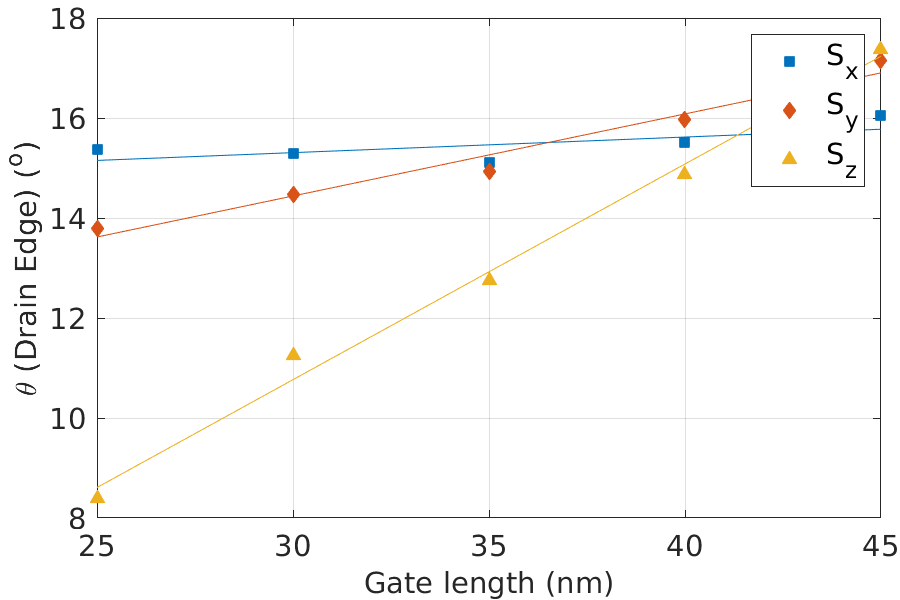}} \hfill
\subfloat[Spin modulation at the drain edge.\label{fig:VisGate}] {\includegraphics[width=0.32\textwidth]{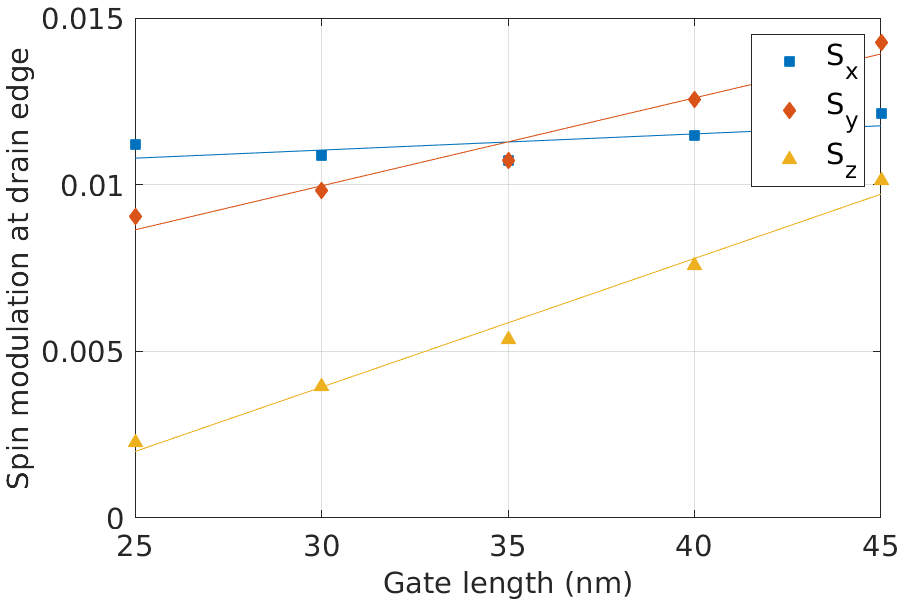}} \\
\subfloat[Minimum spin polarization.\label{fig:MagMinVsGate}]
{\includegraphics[width=0.32\textwidth]{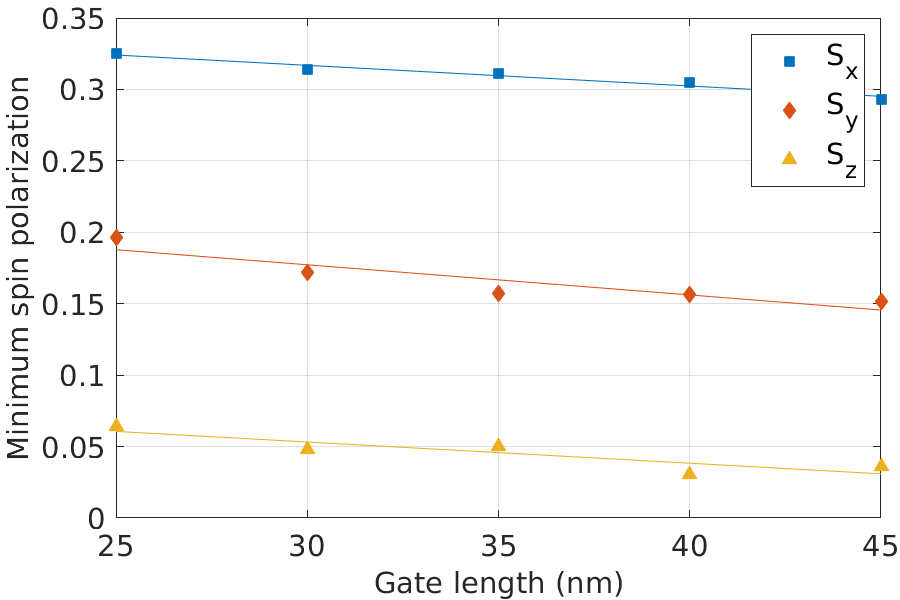}} \hfill
\subfloat[Spin polarization recovery.\label{fig:MagRecVsGate}] {\includegraphics[width=0.32\textwidth]{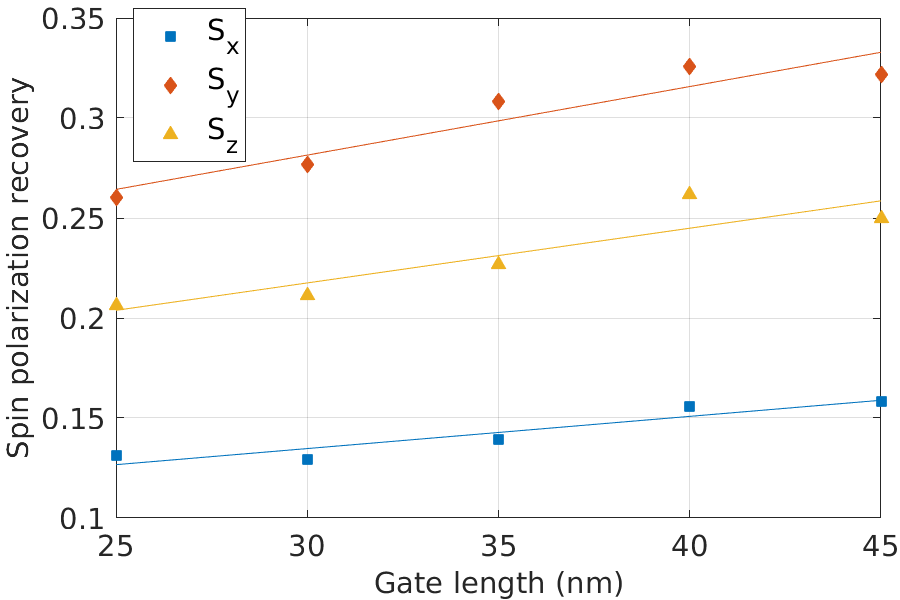}} \hfill
\subfloat[Maximum rotation angle $\theta$.\label{fig:AngMaxVsGate}] {\includegraphics[width=0.32\textwidth]{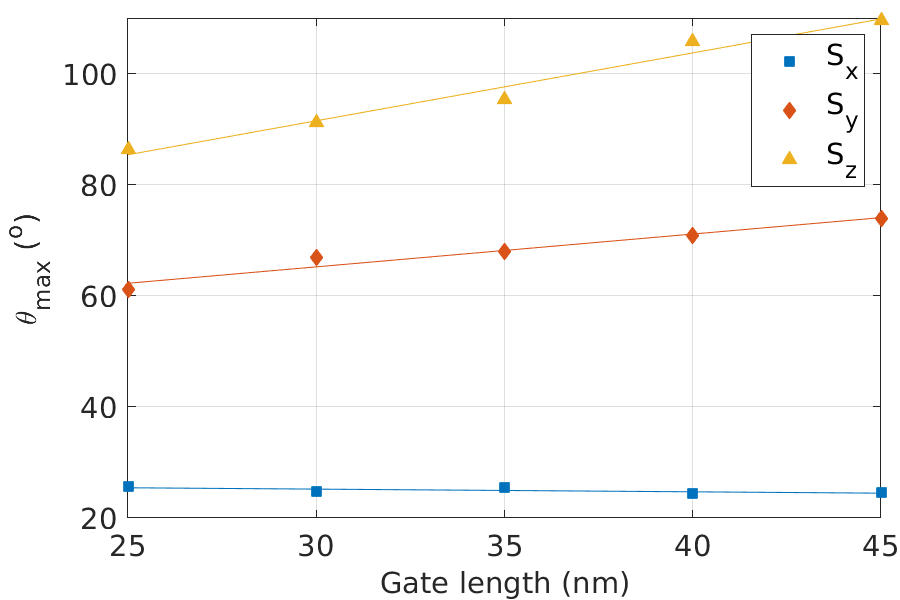}} 
\caption{Spin polarization, rotation angle $\theta$, spin modulation [Eq. (\ref{vis})], all at the drain, then minimum spin polarization, spin polarization recovery, and maximum rotation angle $\theta_{\textrm{max}}$ in $x$, $y$, and $z$ directions (symbols) as a function of the gate length ($V_G$=\SI{0.7}{\volt} and $V_D$=\SI{0.9}{\volt}) with linear fits (full lines) to elucidate trends.}
\label{fig:GateScaling}
\end{figure*}

The gate length is varied from \SI{25}{nm} to \SI{45}{\nano\metre} (see Fig.~\ref{fig:diagram}) while the spacer distances are fixed, i.e., \SI{25}{\nano\metre} for $X_{GG}$ and \SI{26}{\nano\metre} for $X_{GL}$ and $X_{GR}$. Since increasing the gate length increases the length of the channel, the spin polarization at the drain edge should decrease with the increasing gate length. However, Fig.~\ref{fig:MagVsGate} does not show any decrease in the spin polarization at the drain edge with increased gate length. For $S_y$ and $S_z$ initialized spins, the spin polarization at the drain edge slightly increases. This behaviour is clearly evident in the spin polarization along the channel plots in Fig.~\ref{Fig:GateLength} (d-f) and is consistent with earlier simulation results~\cite{Thorpe2017} indicating that the spin polarization does not decrease monotonically along the channel, as it might be expected if the spin polarization decayed exponentially.  Rather, analysis of the spin polarization as a function of the position along the channel suggests that the initial decay is small, resulting in a marginally reduced minimum spin polarization (Fig.~\ref{fig:MagMinVsGate}), but this decrease is more than offset by an increased spin polarization recovery (Fig.~\ref{fig:MagRecVsGate}). The recovery of the spin polarization (akin to a refocusing effect in an NMR system \cite{Hahn1950}) in the gate-to-drain spacer region, where a large fringing field occurs, was reported previously \cite{Thorpe2017}.

At the same time, since the spin-orbit coupling mediated by a large fringing electric field \cite{HanFerry1999} surrounding the region beneath the gate acts over a longer distance, one might expect the polarization angle of the polarization to increase with the increasing gate length, and indeed, both Fig.~\ref{Fig:GateLength}a-c and Fig.~\ref{fig:AngMaxVsGate} show an increase in the maximum polarization angle achieved. Although this increase is followed by a steeper drop between the gate and drain with increasing gate length, Fig.~\ref{fig:thetaVsGate} still shows a linear increase of the polarization angle of the polarization at the left drain edge if the spins are initially $S_y$ and $S_z$ polarized. Combined, the increase in both the spin polarization and the polarization angle at the drain edge with increasing gate length results in an overall spin modulation increase with the gate length for $S_y$ and $S_z$ polarization, as shown in Fig.~\ref{fig:VisGate}.  For $S_x$ polarization, the effect of increasing the gate length on the spin polarization, polarization angle and spin modulation at the drain edge appears to be neutral.

\subsection{Source-To-Gate/Gate-to-Drain Spacer}\label{subsec:Spacers}

The effect, if any, of increasing the source-to-gate spacer length is less obvious to analyse.  A simple model would suggest that increasing the source-to-gate spacer length would be detrimental to the spin polarization at the drain edge due to the increased channel length, and we previously conjectured that increasing the source-to-gate spacer length does not increase the Rashba and Dresselhaus coupling, which leads to the rotation effect and increase in the polarization angle (relative to the initial polarization vector). However, Fig.~\ref{fig:AngMaxVsLS} suggests the maximum polarization angle does in fact increase with increasing source-to-gate spacer lengths.  Furthermore, Fig.~\ref{fig:MagVsLeft} suggests that increasing the source-to-gate spacer length increases the spin polarization at the drain edge, at least for $S_y$ and $S_z$ polarization, while the effect is neutral for $S_x$ polarization.  A similar effect is observed for the polarization angle $\theta$.  Fig.~\ref{fig:AngMaxVsLS} shows that the maximum polarization angle increases in all cases, but the increase is partly offset by a steeper decrease between the gate and drain (Fig.~\ref{Fig:LS}a-c). As a result (Fig.~\ref{fig:thetaVsLeft}) we observe a small net increase in the polarization angle at the drain edge for $S_y$ and $S_z$, while the effect is neutral for $S_x$ polarization. Combining the effect of increasing the source-to-gate spacer length on the spin polarization and polarization angle leads to an increase in the spin modulation at the drain edge for $S_y$ and $S_z$ polarization, but a neutral or marginally detrimental effect for $S_x$ polarization (Fig.~\ref{fig:VisLS}). We also observe a decrease in the minimum spin polarization along the channel (Fig.~\ref{fig:MagMinVsLS}, which is consistent with our expectations that increasing the distance between the source and gate contacts ($X_{GL}$) decreases spin polarization as the electrons spend more time in the source-to-drain region of the channel, thus experiencing more scattering events, before being influenced by the fringing field generated by the gate.  However, the source-to-gate polarization is partially offset by a stronger spin polarization recovery in the gate-to-drain region of the channel (see Figs.~\ref{fig:MagRecVsLS} and \ref{Fig:LS}d-f).

When it comes to increasing the gate-to-drain spacer length, the results are more interesting.  Except for $S_z$ polarization, where the effect is neutral, the minimum spin polarization along the channel decreases (Fig.~\ref{fig:MagMinVsRS}).  However, the decrease is compensated for by an increase in the spin polarization recovery (Fig.~\ref{fig:MagRecVsRS}), because the spin polarization has more time to refocus, resulting in a slight increase in the spin polarization at the drain edge for $S_y$ and $S_z$ polarization, and a neutral effect for $S_x$ (Fig.~\ref{fig:MagVsRight}).  In all cases, we observe that the spin polarization increases or remains constant with increasing the gate-to-drain spacer length.  Again, this is consistent with the refocusing hypothesis and conflicts with a simple model that predict a monotonic decrease in the spin polarization with the channel length \cite{Das2019}. In addition to the increase in the overall spin polarization, Fig.~\ref{fig:AngMaxVsRS} also shows an increase in the maximum polarization angle $\theta_{\max}$ with increasing right spacer length.  However, this increase is more than offset by a sharper drop in the polarization angle between the gate and drain, resulting in a decrease of the polarization angle $\theta$ at the drain edge with the distance between the gate and drain (Fig.~\ref{fig:thetaVsRight} and \ref{Fig:RS}a-c).  For $S_y$ and $S_z$ polarization, the dependence of $\theta_{\max}$ and $\theta$ at the left drain edge on the gate-to-drain length appears to be non-linear, resulting in a similar non-linear dependence of the spin modulation (Fig.~\ref{fig:VisRS}).  In particular, the latter graph suggests that there may be an optimum gate-to-drain spacer length that maximizes the spin modulation for $S_y$ and $S_z$, while increasing the gate-to-drain spacer length appears to be overall detrimental to the spin modulation for $S_x$ polarization.

%%%%%%%%%%%%%%%%%%%%%%%%%%%%%%%%%%%%%%%%
\section{\label{sec:Conclude}Conclusions}
Spin transport in a nanoscale \chem{In_{0.7}Ga_{0.3}As} MOSFET with spin-polarized carriers is explored using ensemble Monte Carlo device simulations.  We report on the behaviour of spin polarization, both its magnitude and polarization angle relative to the initial polarization vector, as a function of temperature ($300$K--$77$K) while varying device lateral geometry.  Specifically, the effects of varying the gate lengths, and the source-to-gate and the gate-to-drain spacers on both the magnitude and the angle of the spin polarization vector are studied. 

The simulation results for the temperature dependence show that the magnitude of the spin polarization in the drain region increases with decreasing temperature. However, decreasing the temperature also reduces the Rashba coupling and the degree of rotation the polarization vector undergoes as the spin-polarized electrons travel through the channel.  Combining these effects show that the spin modulation, which quantifies the modulation of the drain current due to the spin-orbit coupling, actually increases with increasing temperature.  

Our simulation results show that increasing the gate and spacer lengths (within realistic limits) is either neutral or increases the magnitude of the spin polarization at the drain edge due to the refocusing effect~\cite{Thorpe2017}, which is enhanced when the gate or the spacer lengths increase, although this enhancement is limited to the nanoscale dimensions of the device. This increase in the spin polarisation is inconsistent with a simple model, which would suggest that increasing the gate length and, especially, the spacer lengths should decrease the magnitude of the spin polarization at the drain edge due to the increased channel length resulting in a greater depolarization. Beyond that, the simple model predicting a decrease of the spin polarization at the drain edge becomes valid.

While it remains to be seen what architectures are ultimately best for spin-FETs and spintronics applications, the results presented on spin transport and control mechanisms in realistic systems can be utilized to develop new designs for spin applications that properly take into account trade-offs such as increased spin polarization and injection efficiency vs. reduced polarization control at lower temperatures, and optimize controllable design parameters such as channel and spacer lengths to maximize spin-refocusing effects.  Although the simulations are limited to a particular design of an InGaAs FET, spin refocusing effects play an important role in NMR and ESR, and if they could be exploited in semiconductor spinFET designs, this might lead to more practical and functional designs for new applications.

\ack
BT appreciated the support for his studentship, and SS and KK for the research from the S\^{e}r Cymru National Research Network in Advanced Engineering by Welsh Government.

\section*{References}
% -------------BIBLIOGRAPHY-------------------
%\bibliographystyle{apsrev4-1}
\bibliographystyle{iopart-num}
\bibliography{papery.bib}

\providecommand{\newblock}{}
\begin{thebibliography}{10}
\expandafter\ifx\csname url\endcsname\relax
  \def\url#1{{\tt #1}}\fi
\expandafter\ifx\csname urlprefix\endcsname\relax\def\urlprefix{URL }\fi
\providecommand{\eprint}[2][]{\url{#2}}
% Bibliography created with iopart-num v2.1
% /biblio/bibtex/contrib/iopart-num

\bibitem{Awschalom2013}
Awschalom D~D, Bassett L~C, Dzurak A~S, Hu E~L and Petta J~R 2013 {\em
  Science\/} {\bf 339} 1174--1179

\bibitem{Wolf2001}
Wolf S~A, Awschalom D~D, Buhrman R~A, Daughton J~M, von Molin{\'a}r S, Roukes
  M~L, Chtchelkanova A~Y and Treger D~M 2001 {\em Science\/} {\bf 294}
  1488--1495

\bibitem{Datta1990}
Datta S and Das B 1990 {\em Appl. Phys. Lett.\/} {\bf 56} 665--667

\bibitem{IRDS2020}
 2020 International roadmap for devices and systems {(IRDS\texttrademark)},
  2020 edition. \url{https://irds.ieee.org/editions/2020}

\bibitem{Moodera2007}
Moodera J~S, Santos T~S and Nagahama T 2007 {\em J. Phys., Condens. Matter\/}
  {\bf 19}(6) 165202

\bibitem{Rashba1959}
Rashba E~I 1959 {\em Fizika Tverd. Tela\/} {\bf 1} 407--421

\bibitem{Bournel2000}
Bournel A, Dollfus P, Cassan E and Hesto P 2000 {\em Appl. Phys. Lett.\/} {\bf
  77} 2346--48

\bibitem{Bournel2001}
Bournel A, Delmouly V, Dollfus P, Tremblay G and Hesto P 2001 {\em Physica E\/}
  {\bf 10} 86--90

\bibitem{MinShen2004}
Shen M, Saikin S, Cheng M~C and Privman V 2004 {\em Mathematics and Computers
  in Simulation\/} {\bf 65} 351--363

\bibitem{Schliemann2003}
Schliemann J, Egues J~C and Loss D 2003 {\em Phys. Rev. Lett.\/} {\bf 90}(14)
  146801

\bibitem{Cartoixa2003}
Cartoix\'{a} X, Ting Y and Chang Y 2003 {\em Appl. Phys. Lett.\/} {\bf 83}
  1462--1464

\bibitem{Scherubl2016}
Scher\"ubl Z, F\"ul\"op G, Madsen M~H, Nyg\aa{}rd J and Csonka S 2016 {\em
  Phys. Rev. B\/} {\bf 94}(3) 035444

\bibitem{Bandyopadhyay2004}
Bandyopadhyay S and Cahay M 2004 {\em Appl. Phys. Lett.\/} {\bf 85} 1433--1435

\bibitem{TajalliTCS2011}
Tajalli A and Leblebici Y 2011 {\em IEEE Trans. Circuits Syst. I, Reg.
  Papers\/} {\bf 58} 2189--2200

\bibitem{E.Shafir2004}
Shafir E, Shen M and Saikin S 2004 {\em Phys. Rev. B\/} {\bf 70} 241302

\bibitem{Sugahara2016}
Sugahara S, Takamura Y, Shuto Y and Yamamoto S 2016 {\em Field-Effect
  Spin-Transistors\/} (Dordrecht: Springer Netherlands) pp 1243--1279

\bibitem{Nitta:1997}
Nitta J, Akazaki T, Takayanagi H and Enoki T 1997 {\em Phys. Rev. Lett.\/} {\bf
  78}(7) 1335--1338

\bibitem{Engels:1997}
Engels G, Lange J, Sch\"apers T and L\"uth H 1997 {\em Phys. Rev. B\/} {\bf
  55}(4) R1958--R1961

\bibitem{Benbakhi2012}
Benbakhti B, Martinez A, Kalna K, Hellings G, Eneman G, Meyer K~D and Meuris M
  2012 {\em IEEE Trans. Nanotechnol.\/} {\bf 11} 808--817

\bibitem{Kilpi:2021}
Kilpi O~P, Andrić S, Svensson J, Ram M~S, Lind E and Wernersson L~E 2021 {\em
  IEEE Electron Device Letters\/} {\bf 42} 1596--1598

\bibitem{Tomioka:2021}
Tomiok K and Motohisa J 2021 {\em 2021 Silicon Nanoelectronics Workshop
  (SNW)\/} pp 1--2

\bibitem{Lee:2018}
Lee S, Cheng C~W, Sun X, D'Emic C, Miyazoe H, Frank M~M, Lofaro M, Bruley J,
  Hashemi P, Ott J~A, Ando T, Spratt W, Cohen G~M, Lavoie C, Bruce R, Patel J,
  Schmid H, Czornomaz L, Narayanan V, Mo R and Leobandung E 2018 {\em IEDM
  Tech. Dig.\/} pp 912--915

\bibitem{Lu:2018}
Lu W, Lee Y, Murdzek J, Gertsch J, Vardi A, Kong L, George S~M and del Alamo
  J~A 2018 {\em IEDM Tech. Dig.\/} pp 895--898

\bibitem{Thorpe2017}
Thorpe B, Kalna K, Langbein F and Schirmer S 2017 {\em J. Appl. Phys.\/} {\bf
  122} 223903

\bibitem{C.Jacoboni1989}
Jacoboni C and Lugli P 1989 {\em The Monte Carlo Method for Semiconductor
  Device Simulation\/} (Springer Vienna)

\bibitem{Kalna2007}
Kalna K, Wilson J~A, Moran D~A~J, Hill R~J~W, Long A, Droopad R, Passlack M,
  Thayne I~G and Asenov A 2007 {\em IEEE Trans. Nanotechnol.\/} {\bf 6}
  106--112

\bibitem{Kalna2008}
Kalna K, Seoane N, Garc\'{i}a-Loureiro A~J, Thayne I~G and Asenov A 2008 {\em
  IEEE Trans. Electron Devices\/} {\bf 55} 2297--2306

\bibitem{N.Seoane2016}
Seoane N, Aldegunde M, Nagy D, Elmessary M~A, Indalecio G, Garc\'{i}a-Loureiro
  A~J and Kalna K 2016 {\em Semicond. Sci. Technol.\/} {\bf 31} 075005

\bibitem{Ferry2000}
Ferry D~K 2000 {\em Superlatt. Microstruct.\/} {\bf 28} 419 -- 423

\bibitem{AynulIslam2011}
Islam A, Benbakhti B and Kalna K 2011 {\em IEEE Trans. Nanotechnol.\/} {\bf 10}
  1424--1432

\bibitem{Skotnicki2005}
Skotnicki T, Hutchby J, King T~J, Wong H~S and Boeuf F 2005 {\em IEEE Circuits
  Devices Mag.\/} {\bf 21} 16--26

\bibitem{delAlamo2011}
del Alamo J~A 2011 {\em Nature\/} {\bf 479} 317–323

\bibitem{Bournel1997}
Boumel A, Dollfus P, Galdin S, Musalem F~X and Hesto P 1997 {\em Solid State
  Commun.\/} {\bf 104} 85--89

\bibitem{Bournel1998}
Bournel A, Dollfus P, Bruno P and Hesto P 1998 {\em Eur. Phys. J. AP\/} {\bf 4}
  1--4

\bibitem{Oltscher2014}
Oltscher M, Ciorga M, Utz M, Schuh D, Bougeard D and Weiss D 2014 {\em Phys.
  Rev. Lett.\/} {\bf 113} 236602

\bibitem{J.Fabian2007}
Fabian J and Matos-Abiague A 2007 {\em Acta Phys. Slovaca\/} {\bf 57} 565--907

\bibitem{Wang2016}
Wang W and Fu J 2016 {\em Physica B: Cond. Matt.\/} {\bf 482} 14--18

\bibitem{Hubner2009}
Hubner J, Dohrmann S, Hagele D and Oestreich M 2009 {\em Phys. Rev. B\/} {\bf
  79} 193307

\bibitem{Lautenschlager1987}
Lautenschlager P, Garriga M, Logothetidis S and Cardona M 1987 {\em Phys. Rev.
  B\/} {\bf 35} 9174

\bibitem{Kim2012}
Kim T~J, Hwang S~Y, Byun J~S, Barange N~S, Kim J~Y and Kim Y~D 2012 {\em J.
  Korean Phys. Soc.\/} {\bf 61} 97--101

\bibitem{Passler2001}
Passler R 2001 {\em J. Appl. Phys.\/} {\bf 89} 6235

\bibitem{IslamSST2011}
Islam A and Kalna K 2011 {\em Semicond. Sci. Technol.\/} {\bf 26} 055007

\bibitem{KalnaAsenov2002}
Kalna K and Asenov A 2002 {\em Semicond. Sci. Technol.\/} {\bf 17} 579--584

\bibitem{Hahn1950}
Hahn E~L 1950 {\em Phys. Rev.\/} {\bf 80}(4) 580--594

\bibitem{HanFerry1999}
Han J and Ferry D~K 1999 {\em Solid-St. Electron.\/} {\bf 43} 335--341

\bibitem{Das2019}
Das K, Dejene F, Van~Wees B and Vera-Marun I 2019 {\em Appl. Phys. Lett.\/}
  {\bf 114}

\bibitem{Kim2012a}
Kim T~J, Hwang S~Y, Byun J, Diware M~S, Kim J~Y and Kim Y~D 2012 {\em J. Korean
  Phys. Soc.\/} {\bf 61} 1821

\end{thebibliography}
%-----------------APPENDIX--------------------

\onecolumn
\appendix
\section{Parameters used for Temperature dependence of Band Energies\label{AppA}}

Table \ref{temp-param} summarises fitting material parameters for \chem{GaAs}, \chem{InAs}, and \chem{In_{0.3}Ga_{0.7}As} used in the calculations of band energies.
\begin{table*}[t]
\caption{\label{temp-param}Fitting parameters as bandgap $E_{g0}$, $\theta$, and $ \alpha_B$ used to obtain the temperature dependent band energies and Kane parameters. $E_0$ and $E_1$ are the zero and the first energy bands (the zero energy is set to the bottom of the conduction band), $\Delta_0$ and $\Delta_1$ are the zero and the first spin-orbit splitting energies, $P_0$ and $P_1$ are the optical matrix elements for the zero and the first energy bands. $P_0$ and $P_1$ are assumed to be temperature independent for $\chem{InAs}$ and $E_{0/1}+\Delta_{0/1}$ has the same temperature dependence as $E_{0/1}$.}
%\resizebox{\textwidth}{!}{

\centering
\begin{tabular}{c c c c c c c c c c }
Parameter  & \multicolumn{3}{c}{$\chem{GaAs}^a$}& \multicolumn{3}{c}{$\chem{InAs}$}& \multicolumn{3}{c}{\chem{In_{0.3}Ga_{0.7}As}}\\
\multicolumn{10}{c}{\hrulefill}\\
&$E_{g0}(\SI{}{\electronvolt})$&$\alpha_B(\SI{}{\milli\electronvolt})$&$\theta(\SI{}{\kelvin})$&$E_{g0}(\SI{}{\electronvolt})$&$\alpha_B(\SI{}{\milli\electronvolt})$&$\theta(\SI{}{\kelvin})$&$E_{g0}(\SI{}{\electronvolt})$&$\alpha_B(\SI{}{\milli\electronvolt})$&$\theta(\SI{}{\kelvin})$\\
$E_0$&1.571&57&240&$0.414^b$&$28.10^b$&$147^b$&1.224&48.30&212.10\\
$E_1$&4.456&59&323&$4.453^c$&$41.00^c$&$262^c$&4.455&53.60&304.70\\
$E_0+\Delta_0$&1.907&58&240&$0.807^{b,d}$&$28.10^{b,d}$&$147^{b,d}$&1.577&49.03&212.10\\
$E_1+\Delta_1$&4.659&59&323&$4.936^e$&$64.00^e$&$159^e$&4.742&60.50&273.80\\
$2P_0^2/\hbar^2$&30.58&1040&240&-&-&-&-&-&-\\
$2P_1^2/\hbar^2$&8.84&1040&240&-&-&-&-&-&-\\
%\bottomrule
 \end{tabular} \\
 \footnotesize{
a) Ref. \cite{Hubner2009}.
b) Ref. \cite{Passler2001}.
c) Ref. \cite{Kim2012a}.
d) Assumed to have the same dependence as $E_0$ due to lack of data. e) Ref. \cite{Kim2012}
}.
\end{table*}

\section{Auxiliary Figures\label{AppB}}

\begin{figure*}[t] 
\subfloat[$S_x$]{\includegraphics[width=0.32\textwidth]{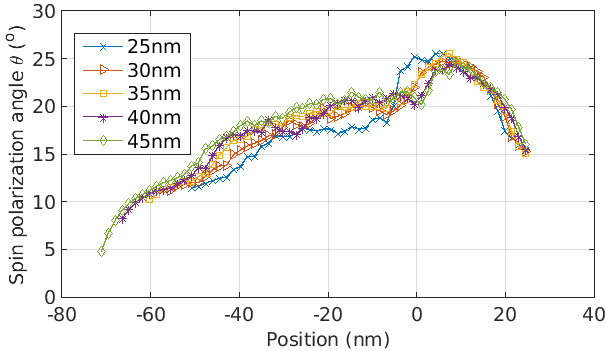}} \hfill
\subfloat[$S_y$]{\includegraphics[width=0.32\textwidth]{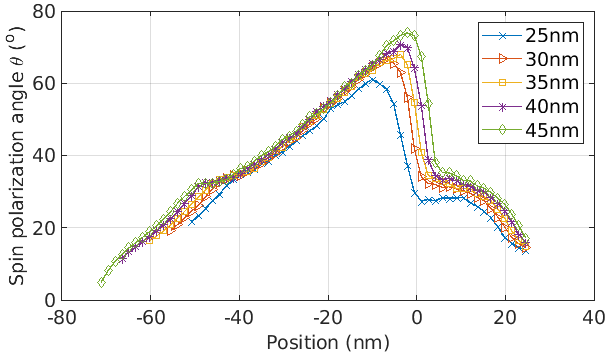}} \hfill
\subfloat[$S_z$]{\includegraphics[width=0.32\textwidth]{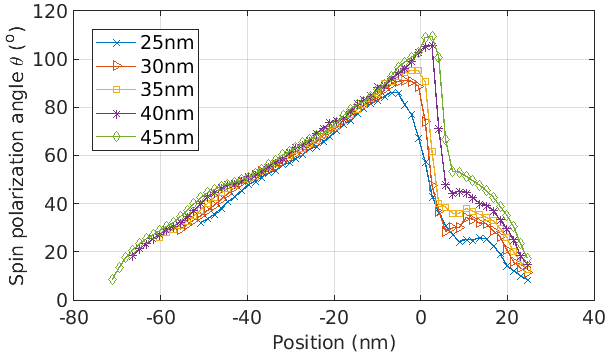}} \\
\subfloat[$S_x$]{\includegraphics[width=0.32\textwidth]{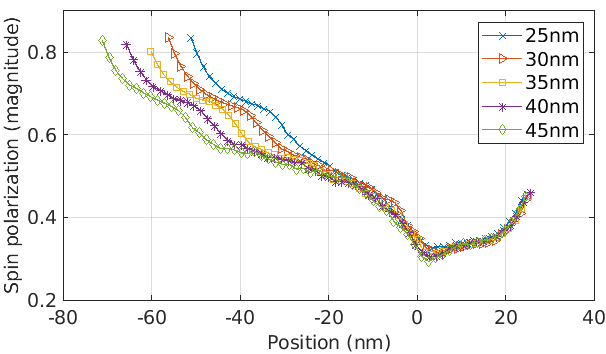}} \hfill
\subfloat[$S_y$]{\includegraphics[width=0.32\textwidth]{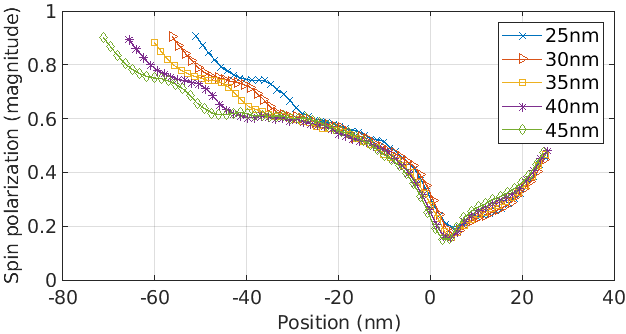}} \hfill
\subfloat[$S_z$]{\includegraphics[width=0.32\textwidth]{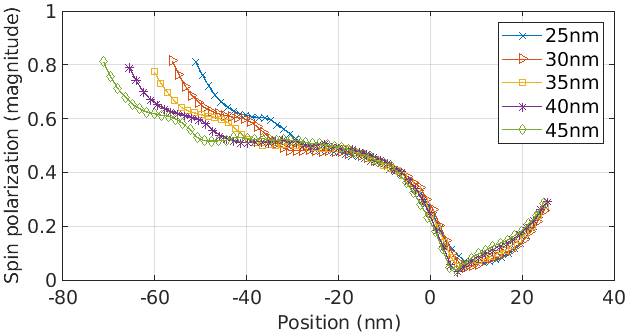}} 
\caption{A spin polarization angle and a magnitude of the spin polarization for the $x$, $y$, and $z$ components as function of the position along the channel for the different gate lengths.}
\label{Fig:GateLength}
%\shipout\box255
\end{figure*}

\begin{figure*}[ht]
\subfloat[$S_x$]{\includegraphics[width=0.32\textwidth]{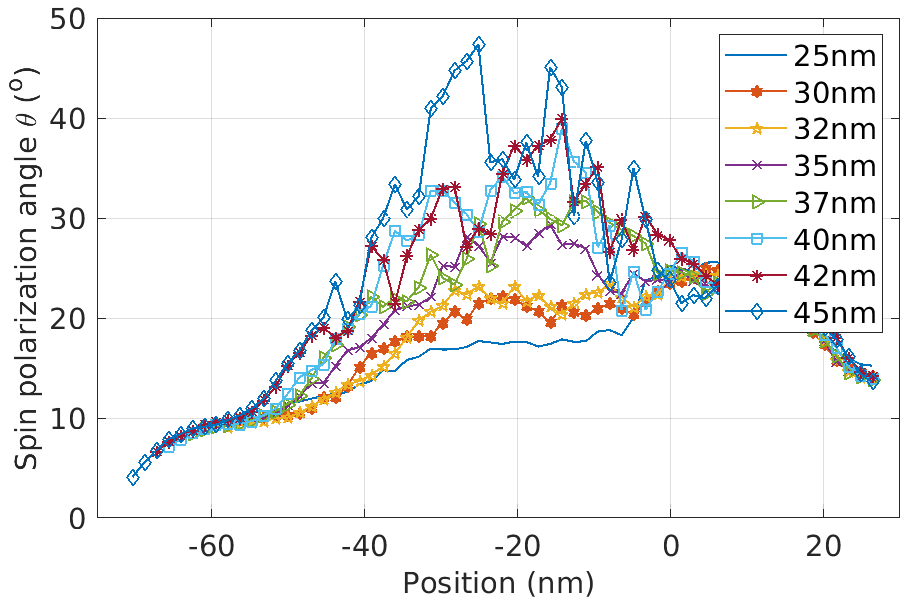}} \hfill
\subfloat[$S_y$]{\includegraphics[width=0.32\textwidth]{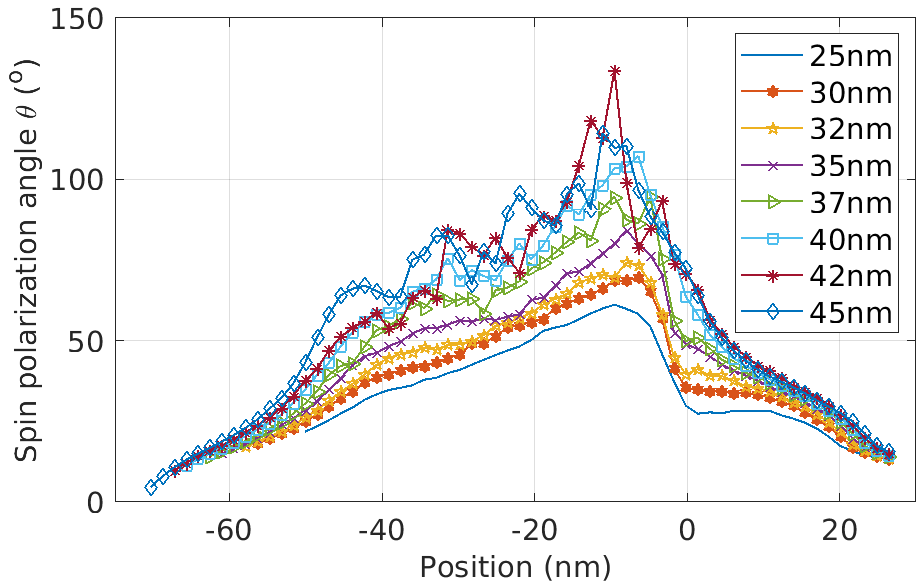}} \hfill
\subfloat[$S_z$]{\includegraphics[width=0.32\textwidth]{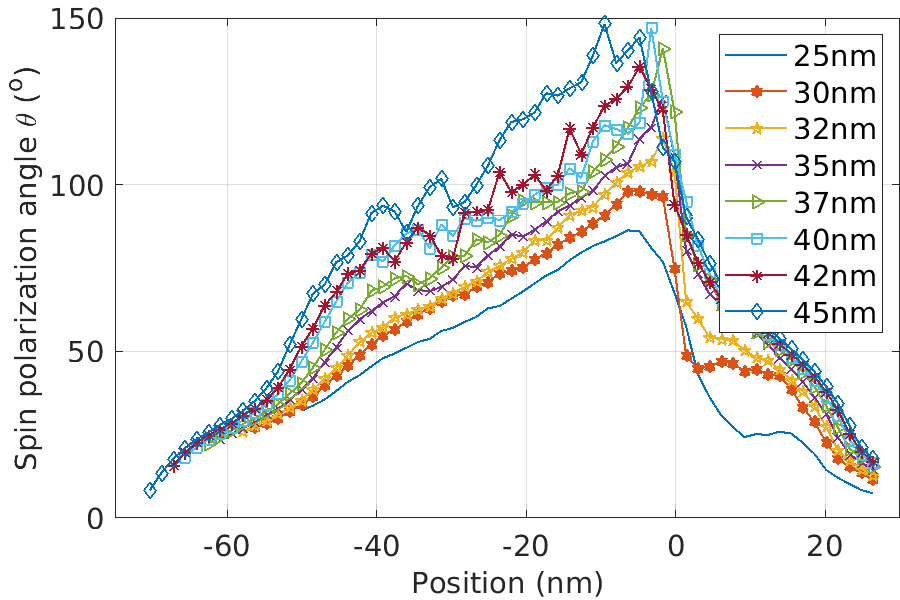}} \\
\subfloat[$S_x$]{\includegraphics[width=0.32\textwidth]{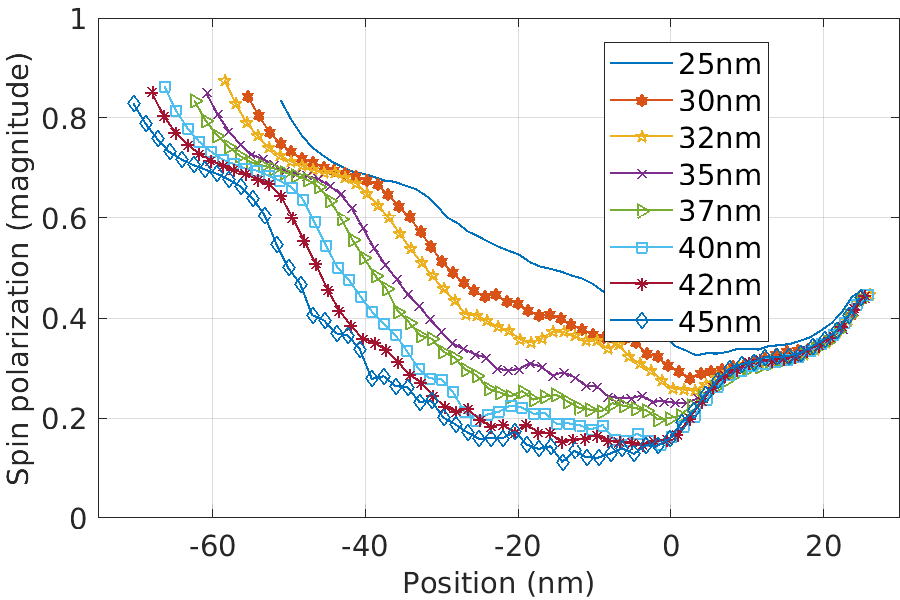}} \hfill
\subfloat[$S_y$]{\includegraphics[width=0.32\textwidth]{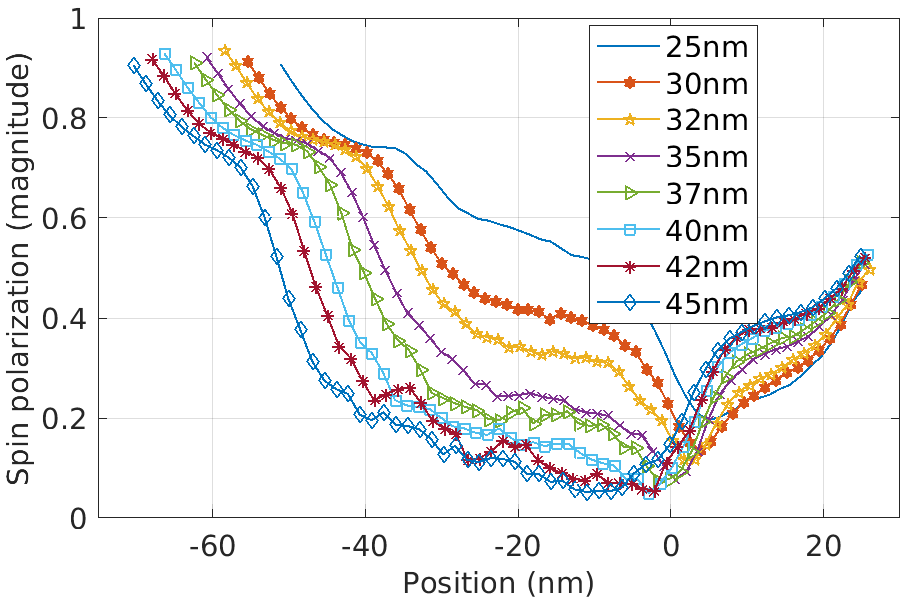}} \hfill
\subfloat[$S_z$]{\includegraphics[width=0.32\textwidth]{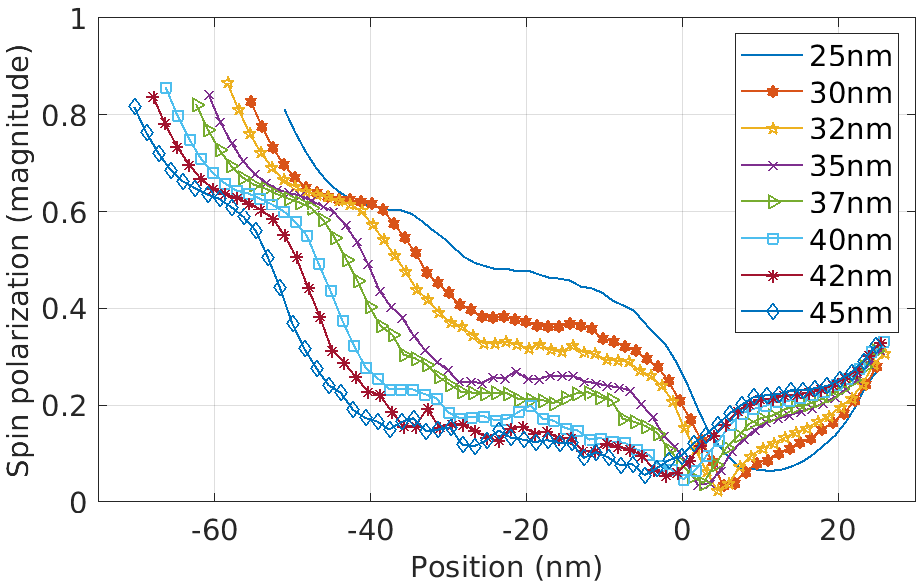}} 
\caption{A spin polarization angle and a magnitude of the spin polarization for the $x$, $y$, and $z$ components as function of the position along the channel for the different source-to-gate spacer lengths.}
\label{Fig:LS}
%\shipout\box255
\end{figure*}

\begin{figure*}[ht]
\subfloat[$S_x$]{\includegraphics[width=0.32\textwidth]{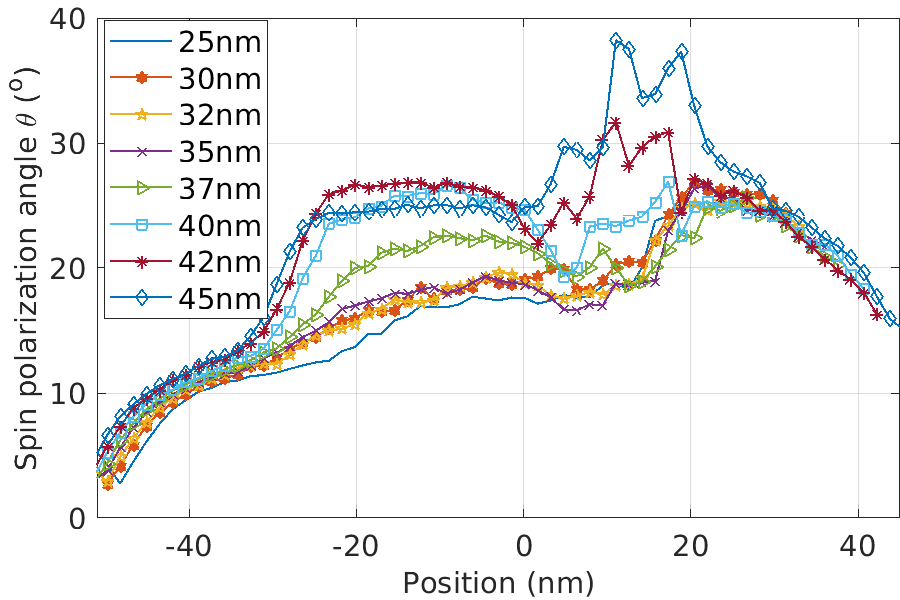}} \hfill
\subfloat[$S_y$]{\includegraphics[width=0.32\textwidth]{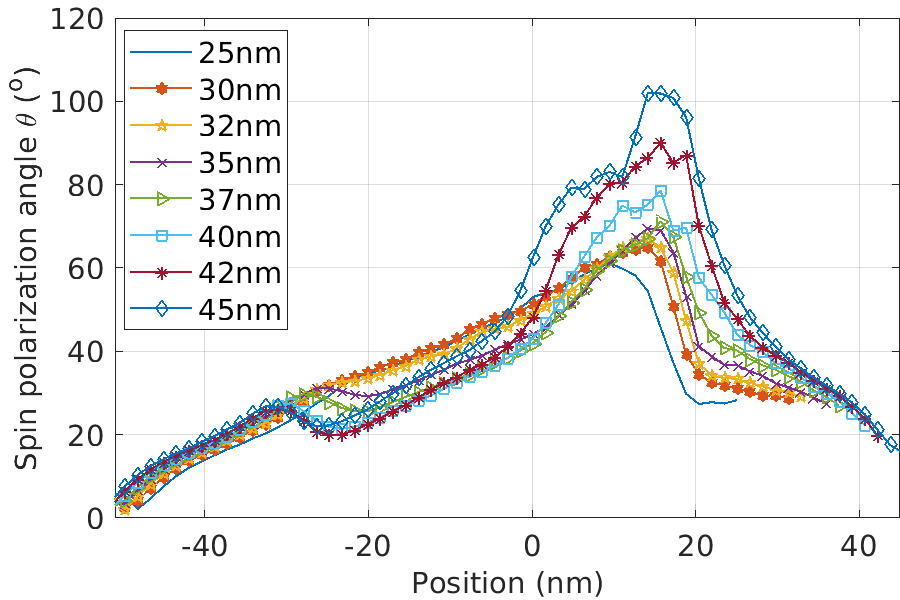}} \hfill
\subfloat[$S_z$]{\includegraphics[width=0.32\textwidth]{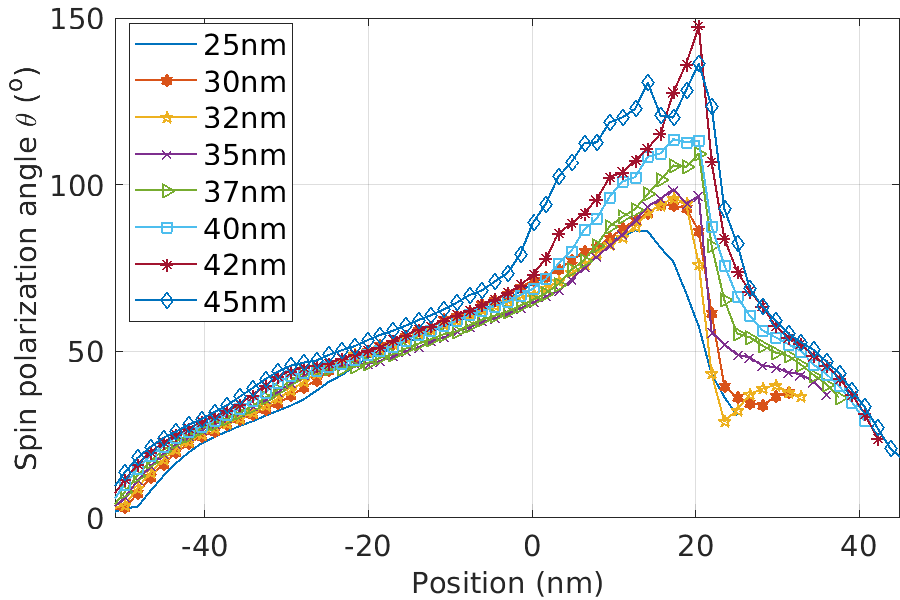}} \\
\subfloat[$S_x$]{\includegraphics[width=0.32\textwidth]{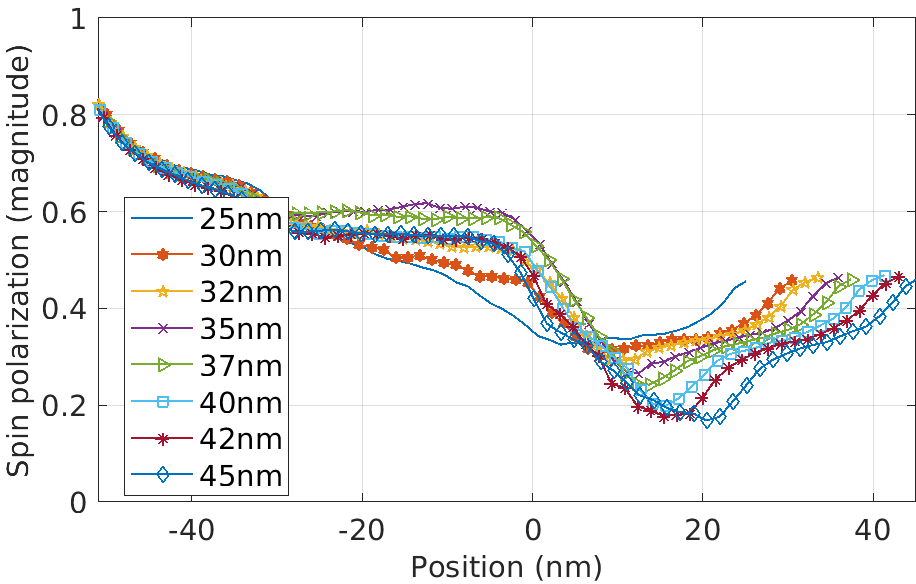}} \hfill
\subfloat[$S_y$]{\includegraphics[width=0.32\textwidth]{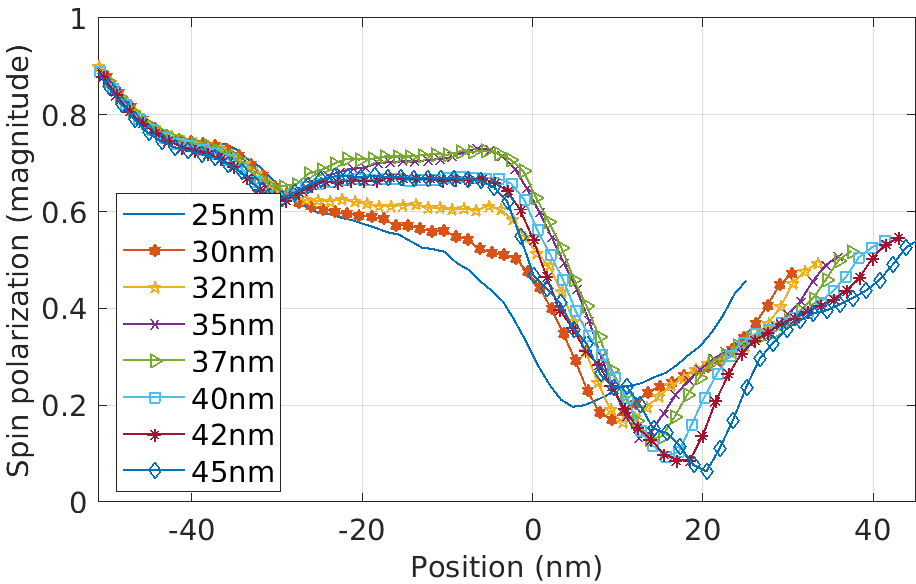}} \hfill
\subfloat[$S_z$]{\includegraphics[width=0.32\textwidth]{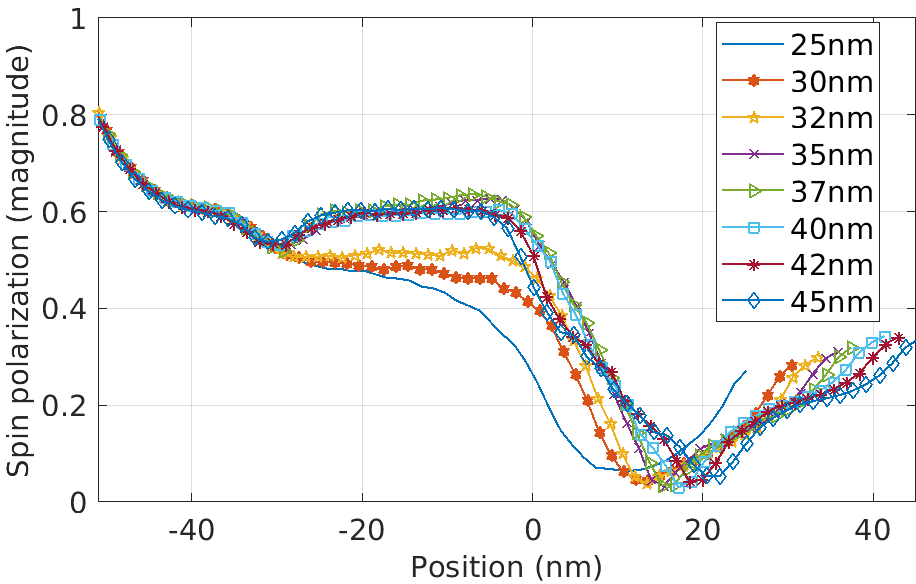}} 
\caption{A spin polarization angle and a magnitude of the spin polarization for the $x$, $y$, and $z$ components as function of the position along the channel for the different gate-to-drain spacer lengths.}
\label{Fig:RS}
%\shipout\box255
\end{figure*}

Figs.~\ref{Fig:GateLength}, \ref{Fig:LS}, and \ref{Fig:RS} plot a spin polarization angle $\theta$ and a magnitude of spin polarization along the device channel at a drive voltage (a gate  voltage of \SI{0.7}{\volt}, a drain  voltage of \SI{0.9}{\volt}) for different gate lengths, source-to-gate spacers, and gate-to-drain spacers, respectively.

\end{document}